\documentclass{aa}  
\usepackage{graphicx}
\usepackage{amsmath,amsfonts,amssymb}
\usepackage{enumerate}
\usepackage{longtable}
\usepackage{epstopdf}
\usepackage{array}
\usepackage{epsfig}
\usepackage{xcolor}
\usepackage[varg]{txfonts}

\usepackage{natbib}
\bibpunct{(}{)}{;}{a}{}{,}

\begin{document} 

\authorrunning{Keszthelyi, Puls, Wade}
\titlerunning{Modeling the early evolution of massive OB stars...}

\title{Modeling the early evolution of massive OB stars with an experimental wind routine}

   \subtitle{The first bi-stability jump and the angular momentum loss problem}

   \author{Z. Keszthelyi\inst{1,2,3},
          J. Puls\inst{1},
          \and
         G. A. Wade\inst{2}}                    
   \institute{LMU Munich, Universit{\"a}tssternwarte, 
   Scheinerstr. 1, D-81679 M{\"u}nchen, Germany
   \and Royal Military College of Canada, 
   PO Box 17000 Station Forces, Kingston, Ontario, Canada, K7K 7B4
   \and Queen's University, Stirling Hall, 
   Kingston, Ontario, Canada, K7L 3N6\\
   \email{zsolt.keszthelyi@rmc.ca}} 
                                                                                        
   \date{Received 3 August 2016 / Accepted 13 October 2016} 

\abstract{Stellar evolution models of massive stars are very sensitive
to the adopted mass-loss scheme. The magnitude and evolution of
mass-loss rates significantly affect the main sequence evolution, and
the properties of post-main sequence objects, including their
rotational velocities.}
{Driven by potential discrepancies between theoretically predicted and
observationally derived mass-loss rates in the OB star range, we aim in particular to investigate the response to mass-loss rates that
are lower than currently adopted, in parallel with the mass-loss
behavior at the ``first'' bi-stability jump.}
{We performed 1D hydrodynamical model calculations of single $20 - 60 \,
\mathrm{M_{\odot}}$ Galactic ($Z = 0.014$) stars where the effects of stellar
winds are already significant in the main sequence phase. We
have developed an experimental wind routine to examine the behavior and
response of the models under the influence of different mass-loss
rates. This observationally guided, simple and flexible wind routine
is not a new mass-loss description but a useful tool based on the
wind-momentum luminosity relation and other scaling relations, and
provides a meaningful base for various tests and comparisons.}
{The main result of this study indicates a dichotomy between solutions of currently debated problems regarding mass-loss rates of hot
massive stars. In a fully diffusive approach, and for commonly
adopted initial rotational velocities, lower mass-loss rates than
theoretically predicted require to invoke an additional source of
angular momentum loss (either due to bi-stability braking, or yet
unidentified) to brake down surface rotational velocities. On the
other hand, a large jump in the mass-loss rates due to the
bi-stability mechanism (a factor of 5 - 7 predicted by Vink et al. (2000, Astronomy \& Astrophysics, 362, 295), but a factor of 10 - 20 in modern models of massive stars) is challenged by observational results, and
might be avoided if the early mass-loss rates agreed with the
theoretically predicted values.}
{We conclude that simultaneously adopting lower mass-loss rates and a
significantly smaller jump in the mass-loss rates over the
bi-stability region (both compared to presently used prescriptions)
would require an additional mechanism for angular momentum loss to be
present in massive stars. Otherwise, the observed rotational
velocities of a large population of B supergiants, that are thought to
be the evolutionary descendants of O stars, would remain unexplained.}


\maketitle 
        
\section{Introduction}
\label{sec:intro}

During their complete evolution, massive stars lose a significant
fraction of their initial mass in the form of stellar winds. This mass
loss has a significant impact on the evolution of massive stars (e.g.,
\citealt{maeder2009}), influencing their properties in two important
ways. First, evidently the actual stellar mass (as a function of time) is
affected by mass loss. The winds of hot OB stars and of their
descendants are sufficiently powerful to remove a significant amount
of mass that their evolutionary paths depend sensitively on the
strength of the wind at the various evolutionary phases. As a
consequence, mass loss (together with rotation, binarity and
metallicity effects) is a key determinant of the final end-states of
massive star evolution. Second, it has recently been noted
\citep{vink2010} that mass loss is also influential for massive star
evolution due to the removal of angular momentum from surface layers
(see also \citealt{langer98}) by the stellar wind. In particular, the
surface rotational velocities and their evolution are determined by
the joint effects of internal transport mechanisms and surface angular
momentum loss due to mass loss. Our main goal in this study is to
investigate the evolutionary implications of currently-debated
uncertainties regarding the magnitude of mass-loss rates of hot
massive stars.

In recent years, it has become clear that the original
assumptions of the radiation-driven wind theory (pioneered by
\citealt{lucy70} and \citealt{cak75}, hereafter CAK) need to be
reconsidered. The discovery of small-scale inhomogeneities in stellar
winds has had a significant impact on the derived mass-loss rates
(e.g., \citealt{Hillier91, Feldmeier03, puls2006, oskinova07,
sundqvist2011, Surlan13}, and summarized by
\citealt{puls2008,sundqvist2013c,puls2015}). Mass-loss rates of hot OB
stars derived both from current X-ray
\citep{cohen2013,leutenegger2013,herve2013,rauw2015}, UV
\citep{sundqvist2011, bouret2012, Surlan13}, and IR \citep{najarro2011}
diagnostics do not agree with the widely-used theoretical rates
derived by \citet{vink2000} and \citet{vink2001}. The apparent
discrepancy is of a factor of between two and three, where rates derived from observations are
lower, when taking the ``Vink-rates'' at face value (see above
references), and typically a factor of two when accounting for
up-to-date abundances in the mass-loss recipe \citep{petrov2016}. Such
changes in the overall mass-loss rates might have severe consequences
for the evolution of massive stars (e.g., \citealt{Meynet1994}). 

Simulating the wind of the famous Luminous Blue Variable, P Cygni (B1
Ia$^+$), \cite{pauldrach90} noted a bi-stable behavior. In their
self-consistent calculations, keeping $T_{\rm eff} = 19.3 $~kK and
varying $\log (L/\mathrm{L_{\odot}}) = 5.74, 5.86, 5.97$, they showed that for
a small increase in the Eddington $\Gamma$, a large impact is seen on
the dynamics of the stellar wind. Instead of a gradual increase,
$\dot{M}$ showed a strong discontinuity versus $\Gamma$, and the
behavior resulted in a jump in $\dot{M}$ accompanied by a
corresponding jump (in the opposite direction) in the terminal velocity.
The triggering mechanism of this bi-stability was attributed to the
behavior of the hydrogen Lyman continuum, namely that it becomes
optically thick at a critical wind density or effective temperature.
The high opacity blocks the flux bluewards of the Lyman edge, and the
metals that have ground state photoionisation edges in this frequency
range shift to a lower ionisation state. The recombination of metals
enhances the radiative line acceleration (since lower ions typically
exhibit more lines), and thus leads to an increase in $\dot{M}$. 

Even if the Lyman continuum were to remain optically thin, a shift in
the ionisation equilibrium could still occur, mainly due to reaching a
critical $T_{\rm eff}$. Regarding this more general situation,
\cite{vink99} identified the dominant role of iron recombination in
the wind. As a massive star evolves and reaches lower effective
temperatures, Fe\,{\sc iv} recombines to Fe\,{\sc iii}, giving rise to
an increased mass-loss rate corresponding to the ``first''
bi-stability jump in the mass-loss rate and terminal velocity of the
wind. 

Calculations by \cite{vink99} confirm the presence of bi-stable
behavior (both with respect to the above theoretical findings and
observational results demonstrating the presence of a jump in the
terminal velocities of B star winds, \citealt{lamers95}), however
their prediction of the jump temperature falls into the range 27.5 -
22.5~kK. This is higher than the \citet{pauldrach90} result, 19.3~kK,
and the values derived by \citet{lamers95} and others (see below),
which indicate a jump temperature of 20.5~kK.

As also noted by \cite{vink99}, a second bi-stability jump should be
expected at around 12~kK, because of the recombination of Fe\,{\sc
iii}. The presence of such a jump has been theoretically
confirmed by \cite{petrov2016}, although  
typically at lower effective temperatures, around 9~kK. 
Interestingly, the model calculations by \cite{petrov2016} predict the first bi-stability to occur at around 20~kK, in agreement
with observations.

Irrespective of the actual position of the jump, the calculations by 
\cite{vink2000} indicate an increase of mass-loss rates by a
factor of between five and seven over the first jump, if the ratio $v_{\infty}/v_{\rm
esc}$ decreases by a factor of two. On the other hand, quantitative
spectroscopy of a sample of Galactic OB supergiants by
\cite{markova2008} provided an (observational) upper limit for any
such increase, namely that the mass-loss rates should at maximum
increase by the same factor as the terminal velocities decrease, meaning
by a factor of between two and three.  However, those results are most consistent with
mass-loss rates that remain constant or even decrease in parallel with
$v_{\infty}$. These and similar findings from other investigations
(e.g., \citealt{cro2006}) have incited vigorous debate about the
behavior of $\dot{M}$ around the expected positions of the
bi-stability jumps (see also \citealt{vink2010}), noting as well that
\cite{lamers95} found a steep decrease of $v_{\infty}/v_{\rm esc}$
over a quite narrow temperature range, while \cite{cro2006} identified
a much more gradual change (see also \citealt{markova2008}). 

\cite{vink2010} argued that a large increase in the mass-loss
rate due to the first bi-stability jump would lead to an efficient
mechanism to brake surface rotational velocities, the so-called
bi-stability braking (BSB). As already outlined, however, the
description of the jump itself is hampered by at least two
uncertainties. Firstly, the observed jump temperature ($T_{\rm eff,
jump1} \approx 20 \, \mathrm{kK}$, \citealt{lamers95, prinja98,
cro2006, markova2008}, based on the behavior of $v_{\infty}$ and/or
$v_{\infty}/v_{\rm esc}$) is much lower than originally considered. \cite{petrov2016} have
also confirmed that improved model calculations yield results quite similar
to observed values. Secondly, the change of the observed mass-loss rate,
that is, the size of the jump at the first bi-stability location (a
factor of 0.4 - 2 when following the analysis of \citealt{markova2008})
does not agree with the \cite{vink99} and \cite{vink2000} values (a
factor of between five and seven). Even more troublesome is that massive star evolutionary
models that adopt the Vink recipe result in an increase of mass-loss
rates by a factor of 10 - 20 at the first theoretical bi-stability
jump location ($\approx 25 \, \mathrm{kK}$)
\citep{br11,ek12,groh2014}.

Since stellar evolution models of massive stars commonly adopt the
Vink recipe, both issues (regarding the overall rates and their
behavior at the bi-stability) might have a fundamental impact on
massive star evolution that has until now not been investigated. Although
we will focus on hot stars, we remind the reader that due to the
adopted position of the second jump, the mass-loss rates of blue
supergiants might also be significantly overpredicted.

Close to the zero age main sequence (ZAMS) of O-type stars, rotational
velocities are relatively high ($200 - 400$ km s$^{-1}$,
e.g., \citealt{howarth97}), while there is overwhelming evidence for a
large population of slowly rotating B supergiants below $20 \,
\mathrm{kK}$ \citep{howarth97,hunter2008,fraser2010,huang2010}. 
Since these B supergiants are thought to be the evolutionary
descendants of rapidly rotating O-type stars, a significant angular
momentum loss should occur during their evolution. 

In this paper, we investigate the impact on massive star evolution
models caused firstly by decreasing the overall mass-loss rates, and secondly,
avoiding a large increase in $\dot{M}$ at the first bi-stability. In
this sense, the angular momentum content of massive stars is considered
to account for observational constraints. In particular, we investigate the model and parameter dependence
inherent to this problem (already noted by \citealt{vink2010}), namely
we evaluate whether there is an actual need for a significant increase
in $\dot M$ around $\approx 20 \, \mathrm{kK}$. 

We will compare the mass-loss rates and surface rotational velocities
resulting from two widely used grids of massive star evolutionary
models that were calculated with different computational codes: the
grids presented by \cite{ek12} and by \cite{br11}. A variety of
comparisons between these two model grids and the underlying codes are
available in the literature
\citep{martins2013,paxton2013,chieffi2013,jones2015}. However, these
works primarily focus on aspects different from those considered in
the present work, which concentrates on the impact of mass loss. Thus,
to independently test the influence of mass-loss rates on massive
stars, we calculate our models by means of the
1D-hydrodynamical code MESA \citep{paxton2011,paxton2013,paxton2015},
after implementing an observationally guided, simple and flexible wind
routine based on the wind-momentum luminosity relation (WLR,
\citealt{kudritzki95,puls96}). This experimental routine - which
includes the possibility of bi-stability jumps - is a powerful tool in
the sense that it can be adjusted to either reproduce observed wind
parameters or to modify these parameters in a simple way. We underscore
that this routine is by no means a new wind model, and will not be
suitable (at least in its present form) for actual production runs. 

This paper is structured as follows. Section 2 introduces the stellar
evolution models/codes considered in the following, while in Section 3
we present our first test with the MESA code. In Section 4 we describe
our experimental wind tool used within our own calculations. In
Section 5 we present the outcome of various model calculations, and
discuss the role of the bi-stability jump. In Section 6 we show that
reduced mass-loss rates (compared to the Vink rates) and the
avoidance of a large jump in $\dot M$ at the bi-stability location
cannot be present simultaneously. In Section 7 we summarize our findings and address relevant
issues that require forthcoming observational tests.

\section{Stellar evolution model}

To perform our calculations, we adopt a widely used, rapidly
developing 1D hydrodynamical stellar evolution code, Modules for
Experiments in Stellar Astrophysics (MESA,
\citealt{paxton2011,paxton2013,paxton2015}) version
r6794\footnote{MESA is free, open source software, available for
download from: \url{http://mesa.sourceforge.net}}. MESA has a wide
range of applicability, and our purpose was to explore the physics and
the evolution of single massive stars through experiments with the
mass-loss rates.

In our comparisons we refer to two often-cited grids of evolutionary
models of massive stars, those of \cite{ek12} using the Geneva code
(GENEC, \citealt{eggenberger2008}), and those of \cite{br11} using the
Bonn evolution code (STERN, \citealt{langer88, petrovic2005}).
Although these grids span extensive ranges in mass and rotational
velocity, in our investigation we focus on 20 - 60 $\mathrm{M_{\odot}}$ models,
either with no rotation or with an equatorial surface rotational
velocity $v_{\rm rot} $ = $300 \, \mathrm{kms^{-1}}$. 

MESA is similar to, and to some extent modeled upon, the Bonn code
\citep{paxton2013}. The bi-stability braking mechanism 
suggested by \cite{vink2010} was based on models similar to those
published by \cite{br11}. For this reason, we adopt a similar
parameter setup (see Table \ref{tab:tab1}). In the following we will
comment on important details and on the major differences from
the \cite{ek12} models. 

\subsection{Abundances}

Our models have been calculated for a metallicity of $Z = 0.014$, using the \cite{asplund2005} mixture of elements, and
adopting the \cite{lodders2003} isotopic ratios. This choice has been
made so that our comparisons to other models (see below) would not
suffer from large differences (as would be the case with the
default $Z = 0.020$ in MESA, or very different mixtures). For
simplicity, we refer to this particular choice as ``Galactic metallicity'' considering that it provides a good description of
massive stars in the solar neighborhood. 

\cite{br11} used $Z = 0.0088$ for the chemical evolution of their
Galactic metallicity models, when individual compositions are
required (e.g., for surface element enrichment). The detailed
metallicity mixture is described by \cite{br11}, supplemented with the
\cite{asplund2005} values for some other elements. The resulting
metallicity is lower than found in other studies, as a consequence of
tailoring the adopted individual elemental abundances. Isotopic
ratios were taken from \cite{lodders2003}. When the opacity tables
(from \citealt{iglesias96}) are required, the \cite{grevesse96}
mixture of elements was adopted and tailored for $Z = 0.014$. (For
different metallicities this is scaled by the iron abundance.)
\cite{ek12} used the \cite{asplund2005} mixture of elements, except for
a different Ne abundance, taken from \cite{cunha2006}. This particular
mixture was then scaled to $Z = 0.014$. Again, isotopic ratios are
from \cite{lodders2003}. The opacities were then generated for this
particular mixture of elements.

\subsection{Convective core overshooting}

In our MESA models, the convective core boundary is determined by the
Ledoux criterion\footnote{In chemically homogeneous layers with
$\nabla_{\mu} = 0$, the Ledoux criterion is equal to the Schwarzschild
criterion.}, 
\begin{equation}
 \nabla_{\rm rad} <  \nabla_{\rm ad} + \frac{\phi}{\delta} \nabla_{\mu}
,\end{equation}
where the nablas are the radiative, adiabatic and chemical gradients,
that is, $\left( \frac{\partial ln T}{\partial ln P} \right)_{\rm ad}$,
$\left( \frac{\partial ln T}{\partial ln P} \right)_{\rm rad}$, and
$\left( \frac{\partial ln \mu}{\partial ln P} \right)$. $\phi$, and
$\delta$ are derivatives from the equation of state, denoting $-
\left( \frac{\partial ln \rho}{\partial ln T} \right)_{\rm P}$, and
$\left( \frac{\partial ln \rho}{\partial ln \mu} \right)_{\rm P,T}$,
respectively \citep{kippenhahn2012}. We have adopted a step-overshooting parameter of $\alpha_{\rm ov} = 0.335$
consistent with the \citet{br11} models. Overshooting is a sensitive
parameter which can significantly modify the outcome of model
calculations since it directly affects the MS
lifetime, as it is well known from evolutionary models
\citep{langer86,schaller92,br11,ek12,chieffi2013,castro2014}.
This issue might deserve an extended discussion; here we provide only some brief comments. 

Currently two methods are adopted to treat convective overshooting in
stellar evolution models: the step method, and the exponential or diffusive method. Both model grids from \citet{br11} and \citet{ek12} use a step
overshooting which refers to an extension of the convective core
$l_{\rm ov}$ by a fraction $\alpha_{\rm ov}$ of the local pressure
scale height $H_{\rm P}$, 
\begin{equation}
l_{\rm ov} = \alpha_{\rm ov} H_{\rm P}.
\end{equation} 
Other studies (mostly of individual stars, e.g.,
\citealt{moravveji2015}) use the exponential method based on
\cite{herwig2000}, which accounts for the change in diffusive mixing
using an additional diffusion coefficient,
\begin{equation}
D_{\rm ov} = D_{0} \mathrm{exp} \left( \frac{- 2 z}{H_{\rm v}}  \right),
\end{equation}
where $D_{0}$ is the diffusion coefficient at the core boundary, $z$
is the vertical distance from the core boundary, and $H_{\rm v}$ is
the local velocity scale height, defined as the exponential
overshooting parameter times the local pressure scale height:
\begin{equation}
H_{\rm v} = f_{\rm ov} H_{\rm P}.
\end{equation} 
Although our MESA models contain an exponential overshoot parameter, this
is effectively not used, and we rely on the step overshoot method
alone, since we aim to perform a consistent comparison. 

In this context, we note that it has become possible to use
asteroseismological measurements to constrain the overshoot parameter 
for hot, massive stars (e.g., \citealt{aerts2015}). This has been
done, for example for $\theta$ Ophiuchi by \citet{briquet2007}, resulting in
a step overshoot parameter, $\alpha_{ov} = 0.44 \pm 0.07$ (at $T_{\rm eff}~\approx$~22~kK and $\log g \approx 3.95$).

Recently, \cite{moravveji2015} claimed that, based on 
asteroseismological measurements, the exponential method (plus
diffusive mixing in the radiative zone) better reproduces observations
than the step overshoot method. For a B-type dwarf star, they derived an
exponential overshoot parameter, $f_{\rm ov}~=~0.016 - 0.017$.

\cite{petermann2015} argued that in stars with strong, observable
magnetic fields, these fields might be sufficiently strong in the deep
interior to suppress core overshooting. \cite{briquet2012} showed
that the observed magnetic star, V2052 Ophiuchi, is reproduced with
models adopting a small overshoot parameter. 
Dynamos operating in the convective core have been proposed to
suppress core overshooting in intermediate-mass stars
\citep{stello2016}. It is reasonable to speculate that this is also
the case for high-mass stars.

It should be also pointed out that even though stellar models with a
calibrated value of overshooting
\citep{moravveji2015,mcevoy2015,castro2014} can reproduce observed
stellar properties, in 1D models the implementation of the physical
problem (small-scale convective motions) is challenging 
\citep{arnett2009}. We also note that \cite{kohler2015} argue that in very
massive stars ($> 60 \, \mathrm{M_{\odot}}$) the value of overshooting is of less
significance since the size of the convective core is large enough
that possible extensions are not relevant.

At a specific mass ($16 \, \mathrm{M_{\odot}}$), the effective
temperature where the models reach the terminal age main sequence
(TAMS) coincides with the effective temperature at which the
rotational velocities are observed to drop significantly
\citep{vink2010}. Indeed, this was the criterion for calibrating the
overshoot parameter as applied by \cite{br11} (based on
\citealt{hunter2008}). This calibration was obtained for stars in the Large Magellanic Cloud (LMC),
and also adopted for Galactic conditions without modifications.
\cite{ek12} determined the overshooting parameter based on the
observed width of the main sequence using models with lower initial
masses, in the range from 1.35 to $9 \, \mathrm{M_{\odot}}$. If the overshoot parameter was smaller than the value $\alpha_{\rm ov} =
0.335$ which is used in our models (e.g., $\alpha_{\rm ov} = 0.1$ as
adopted by \citealt{ek12}, or the corresponding value using
exponential overshooting), and thus the models reached the end of the
main sequence at higher $T_{\rm eff}$, then our quantitative results
would need to be reconsidered. However, our qualitative picture does
not depend on this issue.

\subsection{Rotation, mixing, and magnetic fields}

In MESA (and STERN), the effects of rotation are considered in a fully
diffusive approach. The inclusion of rotation in stellar evolution
models is critical for mixing chemical elements and angular momentum
transport. It has been argued that the implementation of meridional
(Eddington-Sweet) circulation requires an advective treatment
\citep{maeder2009}. Most importantly, the choice of advective or
diffusive approach for the Eddington-Sweet circulation leads to a
qualitatively different behavior of the evolution of the surface
rotational velocities. Moreover, there is also a difference in
calculating the meridional circulation velocity. In MESA and STERN, Eq.~35 from \cite{heger2000} is used, while in the Geneva
code, Eq.~4.38 from \cite{maeder98} is adopted. This can also
affect the way chemical elements and angular momentum are transported
in the stellar interiors.

\cite{ek12} did not include any effects of magnetic fields, while
\cite{br11} considered angular momentum transport due to a
Spruit-Tayler (ST) dynamo \citep{tayler73, spruit2002}. Although
simulations by \cite{braithwaite2006} were reported to produce a
closed dynamo loop using the Tayler instability, \cite{zahn2007} were
unable to obtain a closed loop. As a consequence, the existence of the
Spruit-Tayler dynamo is heavily debated \citep{ruediger2012,
neiner2015}. Even more problematic is that if magnetic fields were
present throughout the radiative zone, a possible interaction with
large scale meridional currents might occur and the combined effects
would need to be considered \citep{maeder2009}. On the other hand,
such effects might justify the use of a fully diffusive treatment
\citep{song2016}.

The \cite{ek12} grid was computed for an initial ratio of ${v_{\rm
rot}{\rm (init)}}/{v_{\rm crit}} = 0.4 $ where the critical velocity
for an Eddington $\Gamma < 0.639$ is $v_{\rm crit} =
\sqrt{\frac{GM}{R_{\rm eq}}}$. The choice of the assumed initial ratio
is based on the peak of the observed rotational velocity distribution
of B-type stars from \cite{huang2010}. \cite{georgy2013} have shown
that this choice reproduces well the observed surface nitrogen
enrichment. 

On the other hand, the \cite{br11} grid was calculated for a wide
range of rotational velocities. To reproduce the observed
nitrogen enrichment from \cite{hunter2008}, the mixing efficiency
parameters from \cite{heger2000} were adopted, and calibrated to $f_c
= 0.0228$ and $f_{\mu} = 0.1$, respectively. While $f_c$ accounts for
the contribution of the rotationally-induced instabilities to the
total diffusion coefficient, $f_{\mu}$ relates to the inhibiting
effect of chemical gradients on the efficiency of rotational mixing
processes. Since MESA follows the Bonn code implementations, after
several tests (see also \citealt{chieffi2013}) we also adopted these
values for the sake of consistency, keeping in mind that these
parameters introduce a considerable uncertainty, though most likely
will not modify our final conclusions.  We also note that the
calibration of mixing efficiencies should depend on initial mass,
initial rotational velocity and metallicity
\citep{demink2009,ek12,georgy2013}. Thus far, it has not been
justified why, for example, the \cite{br11} mixing efficiencies calibrated
for a LMC composition have been also used for their Galactic and SMC
model grids. Furthermore, it must be noted that the calibration of the
rotational mixing efficiency is not independent of the size of the
convective core, hence of the adopted overshoot parameter. Finally,
any adjustment of mixing efficiencies will have an impact on the
angular momentum transport.

\subsection{Mass-loss rates}

We have specifically investigated models that adopt the Vink et al.
prescription, and we have also adopted our experimental wind routine (see
Section \ref{sec:muc}). Besides the actual treatment of the mass-loss
rates (which will be discussed in detail later in the paper), two
major factors deserve special attention.  

\subsubsection{Metallicity dependence}

The \cite{br11} models include a scaling of $\dot M$ with respect to
the surface iron abundance. Instead of an overall metallicity
dependence (e.g., as present in the Vink recipe and adopted by
\citealt{ek12} and in the MESA models), they use a scaling:
\begin{equation}
 \dot M \propto ({\rm Fe_{surf}} / {\rm Fe}_{\odot})^{0.85}
,\end{equation}
where (for reasons of consistency) Brott et al. adopted a value of
7.50 (in units of $\log ({\rm Fe/H}) + 12$ when using number
densities) for the solar iron abundance, following \cite{grevesse96}.
Although metallicity effects are of major interest in a more general
context, we have restricted our investigations to a Galactic
environment. Investigation of other environments require additional
studies, due to the large impact of metallicity on the mass-loss
rates.

\subsubsection{Dependence on rotation} 
\label{sec:rotmdot}

In most cases, stellar rotation has a minor influence on the winds of
massive O-type stars, since for typical rotation rates (far from the
critical value) the centrifugal forces are low, and the distortion of
the stellar shape is insignificant. In extreme cases, two limits
become decisive. The so-called $\Omega$-limit is reached at critical
rotation (at which point the gravitational and centrifugal forces are
equal), while the Eddington limit is reached when $\Gamma = 1$, that is, 
when the luminosity is equal to the Eddington luminosity ($L = L_{\rm
Edd}$). \cite{maeder2000b} combine these limits as the $\Omega
\Gamma$-limit which is reached when the total acceleration (at the
surface) becomes zero, that is, $g_{\rm grav} + g_{\rm cent} + g_{\rm
rad} = 0$. Before proceeding any further, we must address a basic problem
which illustrates another difference between the Geneva and the Bonn
models. 

The definition of the critical velocity by the Bonn group (e.g.,
\citealt{langer98}) and in the MESA models is
\begin{equation}
\label{eq:vcrit_b}
 v_{\rm crit}^{\rm Bonn} = \sqrt{\frac{G M }{R} (1 - \Gamma)}
.\end{equation}
On the other hand, \cite{maeder2000b} pointed out that
Eq.~\ref{eq:vcrit_b} is only valid if the surface radiative flux has a
uniform value (and the surface is not distorted). This is in
contradiction with von Zeipel's theorem for rotating stars (for a
generalization of the von Zeipel theorem for shellular rotation, see
\citealt{Maeder99}, and for an alternative approach see
\citealt{espinosa2011}). If $\Gamma$ is well below a specific
critical value, explicitly calculated as $\Gamma < 0.639$ by
\cite{maeder2000b}, the critical velocity can be calculated
independent of the Eddington $\Gamma$, 
\begin{equation}\label{eq:vcrit_g}
 v_{\rm crit,1}^{\rm Geneva} = \sqrt{\frac{2}{3}\frac{G M }{R_{\rm
 pol}}},
\end{equation}
where at critical rotation $R_{\rm eq} = \frac{3}{2} R_{\rm pol}$. 
For $\Gamma > 0.639$, a modified critical velocity needs to be
defined, which includes the $\Gamma$-dependence.

Now, let us introduce the corresponding scaling factors which are
applied in the evolutionary codes to correct $\dot M$ for
rotational effects. \cite{langer98} (and STERN) use the factor
\begin{equation}
\label{eq:rotf_b}
\frac{\dot{M}(v_{\rm rot})}{\dot{M}(v_{\rm rot} = 0)} = 
\left( \frac{1}{1 - \frac{v_{\rm rot}}{v_{\rm crit}}} \right)^{\xi}
,\end{equation}
with $\xi = 0.43$ (following a fit to the results of
\citealt{friend86} performed by \citealt{Bjorkman93}) and $v_{\rm
crit} = v_{\rm crit}^{\rm Bonn}$.  This formulation gives a reasonable
agreement with the alternative calculations of \cite{pauldrach86} when
the actual velocity is far from the critical one, and the enhancement
of the mass-loss rate due to rotation is on the order of 30 \%.
Equation~\ref{eq:rotf_b} has also been adopted in MESA where $\xi$ is an
adjustable free parameter, and the right-hand side of this equation is
referred to as the rotational $\dot{M}$ boost \citep{paxton2013}.
We note that both \cite{friend86} and \cite{pauldrach86} considered a
sort of maximum rotational effect, by accounting only for particles in
the equatorial plane. A physically motivated alternative to
Eq.~\ref{eq:rotf_b}, using the same assumptions, would result in a
rotational $\dot{M}$ boost of $(1-(v_{\rm rot}/v_{\rm crit})^2)^{1 -
1/\alpha'}$, where $\alpha' < 1$ is related to the force multiplier
parameters of the modified CAK theory (see Eq.~\ref{eq:alf}; cf.
\citealt{puls2008} and references therein). The major point, however,
is that gravity darkening is not considered in this simplified
approach. 

\cite{maeder2000b} include the effects of gravity darkening and derive
the enhancement factor due to rotation as
\begin{equation}\label{eq:rotf_g}
\frac{\dot{M}(\Omega)}{\dot{M} (\Omega = 0)} = 
\frac{(1- \Gamma_{\rm e})^{\frac{1}{\alpha}-1}}
{\left[ 1 - \frac{\Omega^2}{2 \pi G \rho_m} - \Gamma_{\rm e}
\right]^{\frac{1}{\alpha} - 1}},
\end{equation}
where $\alpha < 1$ is the corresponding force multiplier parameter (to
be replaced by $\alpha'$ if ionization effects are accounted for), 
$\Gamma_{\rm e}$ is the Eddington factor for electron scattering
opacity in a a non-rotating star, $\Omega$ is the angular velocity,
$\rho_m$ is the average density of the star, and the term in the
denominator can be approximated by
\begin{equation}
 \frac{\Omega^2}{2 \pi G \rho_m} \, \approx \, \frac{4}{9}
 \frac{v_{\rm rot}^2}{v_{\rm crit}^{2}} 
,\end{equation}
with $v_{\rm crit} = v_{\rm crit,1}^{\rm Geneva}$. The latter
relations are used in GENEC. 
Finally, we note that in contrast to all of the relations above
which result in an increase of mass-loss close to the $\Omega$ or
$\Omega\Gamma$ limit, \cite{mueller2014} suggested alternative
rotating wind models that imply a potential decrease of the total
mass-loss rate, at least for specific models.

\begin{table*}
\caption{Parameter setup in the evolutionary grids and models
discussed in this work.}
\centering
\begin{tabular}{c|c|c|c}
\hline\hline
& \cite{ek12} & \cite{br11} & This study \\
Code & GENEC & STERN & MESA  \\
\hline 
Initial metallicity & 0.014 & 0.0088 & 0.014 \\

$\alpha_{\rm MLT}$ $^1$ & 1.6 / 1.0 & 1.5 & 1.5  \\

Core boundary & Schwarzschild & Ledoux &  Schwarzschild \\
&  &  & /Ledoux \\
 
Overshooting & $\alpha_{\rm ov} = 0.1$ & $\alpha_{\rm ov} = 0.335$ & $f_{\rm ov}$ and $\alpha_{\rm ov}$  \\

Semiconvection & - &  $\alpha_{\rm semi} = 1$  & optional  \\

Radiative opacity$^2$ & OPAL & OPAL & OPAL \\
 
Reaction network$^3$ & NACRE & `own' & REACLIB \\
 
Angular momentum  & advective-diffusive & diffusive & diffusive \\
transport & & \\

Convective mixing & instantaneous & diffusive & diffusive \\
 
Chemical mixing & diffusive & diffusive & diffusive \\
 
Internal magnetic  & - & Spruit-Tayler & Spruit-Tayler/ \\
 field$^4$ & & & none \\
 
Mass-loss rates$^5$ & Vink & Vink & Vink/experimental  \\
\hline
\end{tabular}
\tablefoot{$^1$ The default value used in this work ($\alpha_{\rm MLT} = 1.5$)
might have been changed for specific comparisons (the MESA default
values are $Z = 0.02$ and $\alpha_{\rm MLT} = 2.0$). The mixing
length parameter in the GENEC models is individually specified for
different mass ranges, $\alpha_{\rm MLT} = 1.6$ and $1.0$ for masses below and above 40 $\mathrm{M_{\odot}}$, respectively. \newline 
$^2$ The opacity tables are based on the radiative opacities
from OPAL (\citealt{iglesias93}, \citealt{iglesias96}), but
also several other sources are used, accounting for (very) low and high
temperature ranges. \newline 
$^3$ The nuclear reaction rates are calculated by using and
extending/complementing the specific databases, e.g., NACRE
(\citealt{angulo99}) for the Geneva models; in MESA, REACLIB
(\citealt{cyburt2010}) or optionally NACRE can be used. For our own
MESA calculations, we used the default `basic.net'. The Bonn code uses `own'
reaction networks, but we were not able to identify the corresponding sources.
\newline 
$^4$ \cite{br11} used the Spruit-Tayler dynamo mechanism for angular
momentum transport but not for chemical mixing. Our own MESA models
have be calculated in analogy. \newline
$^5$ Other mass-loss prescriptions are available for different
evolutionary stages.}
\label{tab:tab1}
\end{table*}

\section{First tests with MESA}
\subsection{Non-rotating models -- comparisons with MESA}

A variety of model grids for massive stars are available in the
literature and online (and numerous comparisons between them, for example,
\citealt{martins2013}, \citealt{paxton2013}, \citealt{chieffi2013},
\citealt{jones2015}). To convince ourselves of the capabilities of
MESA, as a first step we compared or reproduced the main sequence of
non-rotating Galactic ($Z = 0.014$) model grids published by
\citealt{ek12} (using GENEC) and \citealt{br11} (using
STERN). We recall that \cite{br11} also use $Z = 0.014$ for the
opacity calculations in their Galactic models. Although non-rotating
models are somewhat unphysical, it is useful to compare such models
calculated by different codes, since this provides valuable
information independent of the different treatment of rotation in the
various codes. Particularly, clues about the sensitivity of the model
calculations to the input parameters can be obtained, by testing their
influences on the HRD tracks. Note that we did not attempt to compare
the post-MS phases nor the $T_{\rm eff}$ range in which line-driven
winds are no longer applicable. 

\begin{figure} 
\centering
 \resizebox{9cm}{!}{\includegraphics{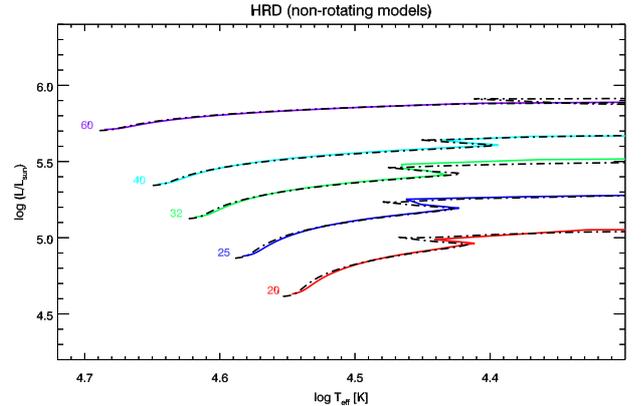}}
\caption{Comparison between non-rotating Galactic Z Geneva
evolutionary tracks on the MS published by \citealt{ek12} (black
dashed), and our MESA models calculated with a similar setup (colored
lines). Initial masses in solar units are indicated next to the
tracks. A step overshoot parameter, $\alpha_{\rm ov} = 0.1$, was
used.}
\label{fig:comp_hrd_g}
\end{figure}

\begin{figure} 
\centering
  \resizebox{9cm}{!}{\includegraphics{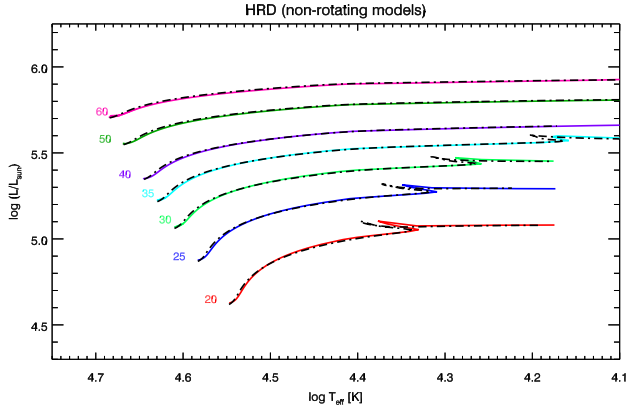}}
\caption{Comparison between non-rotating Galactic $Z$ Bonn
evolutionary tracks on the MS published by \citealt{br11} (black
dashed), and our MESA models calculated with a similar setup (colored
lines). A step overshoot parameter, $\alpha_{\rm ov} = 0.335$, was
used. Stars above 40 $\mathrm{M_{\odot}}$ are still on the
main sequence at rather low effective temperatures. The lack of
hooks at higher masses is due to envelope inflation (N. Langer,
priv.comm.), treated similarly in STERN and MESA.}
\label{fig:comp_hrd_b}
\end{figure}

The overall reproduction of the main sequence as calculated in both
grids is excellent, although small qualitative and quantitative
differences exist (see Figs.~\ref{fig:comp_hrd_g} and
\ref{fig:comp_hrd_b}). One of these issues refers to the starting
point of the ZAMS which occurs at higher effective temperatures in the
Geneva tracks. Moreover, a sensitive point of comparison is the
evolution during and right after core hydrogen exhaustion (the
Terminal Age Main Sequence, TAMS). There are indeed small differences
around the ``hooks'' where the stars first turn back on the HRD
towards increasing $T_{\rm eff}$, and then turn again redward. The
reason for this behavior (in a simplified picture) is that at this
point the star undergoes significant internal structural changes. The
core, which at this point consists almost entirely of helium, starts
to contract on a thermal timescale. This contraction is due to
exceeding the Sch{\"o}nberg-Chandrasekhar limit, which states that the
pressure in the core cannot sustain the weight of the envelope above a
given core mass to total mass ratio ($\frac{M_c}{M} \gtrsim 0.1$). The
core contraction leads to an overall contraction, and an increase of
the core temperature, which maintains the star in hydrostatic
equilibrium. This also results in heating the envelope which will then
initiate hydrogen shell burning in the envelope, which leads to
increasing radius and decreasing surface temperature. Therefore, the
hook in the HRD corresponds to a contracting phase with increasing
$T_{\rm eff}$, and the following redward evolution is due to shell
H-burning. This process, however, is sensitive to the
prescription of convective mixing: using either
instantaneous (GENEC) or diffusive mixing (STERN, MESA) might
partly be responsible for small differences when comparing the
tracks.

To reproduce these models as closely as possible, we used an
identical parameter setup as described in the studies of \cite{ek12}
and \cite{br11}, from which one of the most important is the (step-)
overshooting, set to $\alpha_{\rm ov} = 0.1$ in the former, and
$\alpha_{\rm ov} = 0.335$ in the latter case. Furthermore, for this
comparison we applied the mass-loss prescription of \cite{vink2001},
as was also done in the other two studies. It should be noted,
however, that the implementation of the Vink prescription is not
exactly the same in the three codes. 

To conclude, we have demonstrated that the general behavior of the MS
evolution of massive stars, as reported previously using the Geneva
and Bonn codes, can be reproduced using MESA. However, the following
points must be taken into account. Firstly, this reproduction does not mean that the models agree for all of
 their detailed physical parameters. As an example, the MS lifetimes of
 the models are different. Secondly, the free parameters are not independent, and hence it is possible 
 to obtain similar results with a different parameter setup. For
 example, simultaneously increasing the overshooting and reducing the
 convective mixing efficiency can produce (almost) the same
 evolutionary track. This situation becomes further complicated in
 rotating models. Thirdly, since the codes are not identical, but have certain differences 
 in their implementations, their results will always have at least
 small discrepancies. For instance, the difference of the model
 properties at the initial timestep(s) reflect the parameters of the
 starting model which is loaded. The differences at the TAMS might be
 attributed to more delicate issues, since it is quite challenging to
 simulate the very rapid and very large internal changes that these
 models undergo immediately after core hydrogen exhaustion. The most
 influential free parameter in this respect is overshooting. 

\subsection{Rotating models --  comparison between MESA and the Brott et al. models.}

Rotation plays a crucial role in the evolution of stars: it is
responsible for the mixing of elements and the transport of angular
momentum (e.g., \citealt{langer98, maeder2009}). Moreover, fast
rotation modifies the strength and the topology of hot star winds (see
Sect.~\ref{sec:rotmdot}).  The inclusion of rotation in 1D stellar
evolution codes is a non-trivial task. In a spherically symmetric,
non-rotating case the models are well described by concentric,
spherically-symmetric equipotential layers.  However, if rotation is
present, and shellular rotation \citep{zahn92} is assumed, the
(pseudo-)equipotentials are deformed due to the centrifugal force, and
the distance to the center of two given surface elements on the same
(pseudo-)equipotential is no longer described by a unique radius.
Therefore a one-dimensional treatment of rotation must work with
suitable averages along pseudo-equipotential surfaces. The inclusion
of rotation and its effects requires a special treatment in
evolutionary codes which shows remarkable differences between the two
schools, that is, the Bonn (e.g.,
\citealt{langer98,heger2000,heger2005,petrovic2005,yoon2005,br11}) and
Geneva (e.g.,
\citealt{maeder2000b,maeder2003,hirschi2004,eggenberger2008,ek12})
groups. An attempt to present a unified description of stellar
rotation in 1D evolution codes was provided by \cite{potter2012} using
the ROSE code.
\begin{table}
\caption{Surface rotational velocities in the \cite{br11} grid and our
experimental model grid for four representative initial masses.}
\centering
\begin{tabular}{ccccc} 
\hline\hline
$M_{\rm initial}$ [$\mathrm{M_{\odot}}$] & 20  & 30 & 40  & 60 \\
\hline
STERN &  &   &  & \\
$v_{\rm rot}({\rm init})$ [km s$^{-1}$] & 274 & 269  &  265 & 262 \\
$v_{\rm stable}$ [km s$^{-1}$] & 273  & 268  & 265  & 261 \\
$\Delta t$ [$10^{5}$ yr] & 2.02  & 1.20 & 0.68  & 0.44 \\
\hline
MESA &  &   &  & \\
$v_{\rm rot}({\rm init})$ [km s$^{-1}$] & 274  & 269 &  265  & 262 \\
$v_{\rm stable}$ [km s$^{-1}$] & 277 & 271 &  266  & 260 \\
$\Delta t$ [$10^{5}$ yr] & 3.58  & 0.64  & 0.25  & 0.09\\
\hline
\end{tabular}
\tablefoot{ $v_{\rm rot}({\rm init})$ refers to the input
initial surface rotational velocity, and $v_{\rm stable}$ is the value corresponding to a stable
surface rotational velocity caused by the efficient angular momentum
transport due to a Spruit-Tayler dynamo and the fully diffusive
treatment of angular momentum transport.  $\Delta t$ is the age of the
model at stabilization.}
\label{tab:rot_b}
\end{table}

Since the MESA implementation of rotation follows (but is by no means 
identical to) the Bonn code, a comparison of MESA with this code can
be conveniently performed by tuning the free parameters as described
by \cite{br11} ($f_c = 0.0228$ and $f_{\mu} = 0.1$). Thus we consider
rotationally induced instabilities as diffusive processes (see
\citealt{heger2000,paxton2013}), including the dynamical and secular
shear, the Goldreich-Schubert-Fricke instability, and the
Eddington-Sweet circulation for chemical mixing. For consistency, we
also turn off the Solberg-Hoiland instability for any transport
mechanism (Ines Brott \& Norbert Langer, priv.comm.). In analogy to
the work by \cite{br11}, transport of chemical elements due to the
Spruit-Tayler dynamo is ignored. However, angular momentum transport
due to the Spruit-Tayler dynamo is included. 

After relaxation of the initial models, the rotational velocities
provide an excellent match to the \cite{br11} values. In Table
\ref{tab:rot_b}, we show the initial surface rotational velocities,
noting that besides relaxation effects, the surface $v_{\rm rot}$
remains close to its initial value. We conclude that this
behavior is mainly due to the effects caused by meridional circulation
and the Spruit-Tayler dynamo. As already pointed out, the real
challenge is that magnetic fields and meridional circulation may
interact \citep{maeder2009}. Such potential interaction, however, is
not yet understood and needs to be explored in detail. Thus the
inclusion or exclusion of magnetic fields, and the advective-diffusive
vs. purely diffusive treatment of angular momentum transport lead to
major differences between the models from different authors
\citep{br11,ek12,chieffi2013,paxton2013}. 

\begin{figure}
\centering
  \resizebox{9cm}{!}{\includegraphics{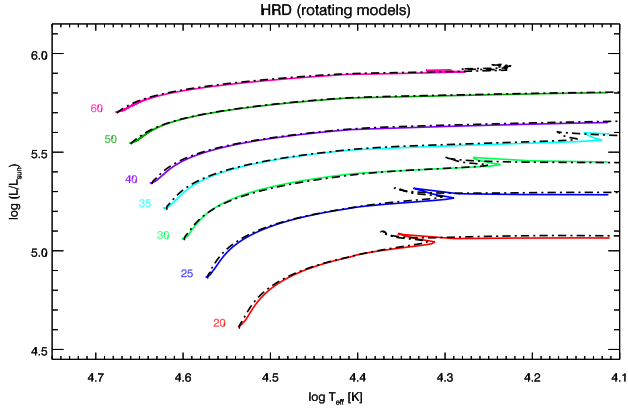}}
\caption{Comparison between rotating Galactic $Z$ Bonn evolutionary
tracks on the MS published by \citealt{br11} (black dashed) and our
MESA models calculated with a similar setup (colored lines).  The
average initial surface rotational velocity of the grid models is
$\approx 270$~km~s$^{-1}$ (exact values indicated in
Table~\ref{tab:rot_b}). A step overshoot parameter, $\alpha_{\rm ov}~=~0.335$, was used.} 
\label{fig:comp_hrd_b_rot}
\end{figure}

In Fig.~\ref{fig:comp_hrd_b_rot}, we compare
our experimental MESA models with the rotating Galactic
models of \cite{br11}, and we can confidently rule out a possible
degenerate solution resulting from interacting input parameters,
since the most important quantities related to rotation ($v_{\rm rot}$
on the MS, diffusion coefficients for mixing and transport processes)
agree extremely well. For further details, we refer to the comparisons
provided by \cite{paxton2013}. 
Finally, we note that due to the similarities between STERN and MESA
and because of the good reproduction of the results of \cite{br11} as
obtained here, we will adopt similar parameters for most of our
standard evolutionary models in the following (see
Table~\ref{tab:tab1}).

\section{An experimental wind routine for hot stars}
\label{sec:muc}

Mass loss has a major impact on the evolution of hot, massive stars
(e.g., \citealt{Meynet1994}), but consequences of uncertainties
related to wind strength and the behavior of (at least) the first
bi-stability jump (see Sect.~\ref{sec:intro}) have not been tested in
stellar evolution model calculations thus far. To this end, we aimed
to develop a simple and flexible tool to reproduce different
mass-loss scenarios. To avoid confusion regarding the different wind schemes in evolutionary
codes, we stress that our new routine can only be applied for hot
($\approx 50 - 15 \, \mathrm{kK}$)\footnote{by specifying
corresponding parameters for the second bi-stability jump, one could
extend this range to $\approx 9,000$~K.} and massive ($ \approx 8 - 60 \,
\mathrm{M_{\odot}}$) main-sequence and post-MS stars. In its current state, it
is in an experimental phase, and not applicable for production runs.

In brief, we implemented a wind routine based on the semi-empirical
WLR, which in turn can be
understood from theoretical scaling relations of mass-loss rate and
terminal velocity \citep{kudritzki95, puls96}. Multiplying the 
wind momentum rate with the square-root of the stellar
radius, the WLR can be written as
\begin{equation}
\dot{M}v_{\infty} (R/R_{\odot})^{1/2} \propto L^{1/\alpha'},
\end{equation}
if $\alpha'$ (see below) is close to 2/3. 
Conveniently, this equation is expressed in logarithmic form:
\begin{equation}
\log D_{\rm mom} = x \, \log L + \log D_0,
\end{equation}
where $x = 1/\alpha'$, and the offset $D_0$ depends on metallicity
and spectral type. In these relations, 
\begin{equation}\label{eq:alf}
\alpha' = \alpha - \delta,
\end{equation}
depends on the force multiplier parameters $\alpha$ and $\delta$
\citep{abbott82,pauldrach86}, related to the radiative line
acceleration
\begin{equation}
g_{\rm rad} \, \propto \, \left( \frac{1}{\rho}
\frac{\mathrm{d}v}{\mathrm{d}r} \right)^{\alpha}  \left( \frac{n_{\rm
e11}}{W} 
\right)^{\delta}.
\end{equation}
Here, $n_{\rm e11}$ is the electron density (in units of
$10^{11}$~cm$^{-3}$), and $W$ the dilution factor. $\alpha$ refers to
the exponent of the line-strength distribution function, and also
provides the ratio between the line force from optically thick lines
and the total one. $\delta$ quantifies changes in the ionization
balance. Since, for typical OB star wind conditions, $\alpha = 0.6 -
0.7$ and $\delta = 0.02 - 0.1$ \citep{puls2000, kudritzki2000}, the
above requirement of $\alpha' \approx 2/3$ is usually fulfilled.

This is, however, not true for the whole spectral range. For example, for
A supergiants, $\alpha' \approx 0.4$ \citep{puls2000}, and according
to \cite{lamers95} the force multiplier parameters and thus $\alpha'$
become discontinuous around $T_{\rm eff} =~21 \, \mathrm{kK}$.
Nevertheless, for simplicity we assume a global $\alpha' = 1/x$ to be
representative for the complete $T_{\rm eff}$ range under
consideration. For future studies, we advise accounting for a
proper temperature dependence, i.e., $\alpha'(T_{\rm eff})$ and $\log
D_0(T_{\rm eff})$. In most cases we have adopted a fixed value of $x =
1.84$ from \cite{mokiem2007a}, which is consistent with theoretical
values in the OB star range (see above).

The wind momentum rate is a very useful quantity, particularly when 
comparing observations with theoretical predictions. This is why many
studies (e.g.,
\citealt{puls96,kudritzki2000,rep2004,mokiem2005,martins2005,mokiem2007b})
have tried to constrain the WLR observationally. In the following, we will
concentrate on Galactic conditions. We recall that the observed WLR in
most cases constrains very well luminosity class I stars, whereas for
other classes this relation may be ambiguous (e.g., the ``weak
wind problem'', see \citealt{puls2008} and references therein, and
also \citealt{hue2012}). Note that to first order, at least the theoretical WLR does not depend on the luminosity class (see
also \citealt{vink2000}).

Different studies have derived different parameters for the WLR. For
comparison, some of these are listed in Table \ref{tab:xd}. Note that
$\alpha'$ and $\log D_0$ correlate strongly with one another (since
they are derived from a linear fit). This becomes obvious when, for example,
comparing corresponding values with and without clumping correction
from the same study. 
\begin{table}
\caption{WLR parameters for Galactic early-type stars.}
\centering
\begin{tabular}{l c c}
\hline\hline
Reference & $x = \frac{1}{\alpha'}$ & $\log D_0$ \\
\hline 
\cite{kudritzki2000}, OI &  1.51 &  20.69\\
\cite{vink2000}, OB & 1.83 & 18.68   \\
\cite{rep2004}, OI, *cl & 2.00 & 17.98 \\
\cite{markova2004}, O, `case D' & 1.90  &  18.58 \\
\cite{martins2005}, O & 3.15  & 10.29 \\
\cite{mokiem2005}, O, early B & 1.86  & 18.71 \\
\cite{mokiem2005}, O, early B, *cl & 1.58  & 20.16 \\
\cite{mokiem2007a}, O, early B & 1.84  & 18.87 \\
\cite{mokiem2007a}, O, early B, *cl & 1.56 & 20.23 \\
\hline
\end{tabular}
\tablefoot{Except for Vink et al., all values have been derived from
observational results. Investigated spectral types are provided. The
theoretical Vink values refer to the hot side of the bi-stability
jump. The values denoted by ``*cl'' have been derived from mass-loss
rates corrected for wind clumping. Units of $D_0$ are in the cgs
system.}
\label{tab:xd}
\end{table} 

In most cases, these observational results confirm the validity of the
WLR concept, although there is significant scatter in the
corresponding coefficients. Moreover, most of these values
overestimate the actual wind momentum rates, since they were derived
for ``smooth'' winds, without correcting for wind inhomogeneities
(e.g., clumping). 

In the following, we outline some details of our experimental wind routine.
The terminal velocity scales with the escape velocity,
\begin{equation}\label{eq:vinf1}
 \frac{v_{\infty}}{v_{\rm esc}} = f_{\rm vinf} \, (\mathrm{spectral \, type}, \mathrm{metallicity}).
\end{equation}
For typical O-star conditions, $f_{\rm vinf} = 2.65$
\citep{kudritzki2000}. We note that in our description $f_{\rm vinf}$ is
an adjustable input parameter that can be calibrated based on
observations
\citep{groe89,lamers95,prinja98,prinja90,cro2006,markova2008}. The
effective escape velocity (accounting for Thomson scattering) is
\begin{equation}\label{eq:vesc1}
 v_{\rm esc} = \left(\frac{2\, G \, M}{R} (1 - \Gamma_{\rm e})  \right)^{1/2}.
\end{equation}
In our formulation (which is consistent with the basic CAK approach),
the Eddington factor needs to be calculated for pure electron
scattering,
\begin{equation}
 \Gamma_{\rm e} = \sigma_{\rm e} \frac{L}{4 \, \pi \, c \,G \, M}  \, .
\end{equation}
Furthermore, we assume hydrogen to be fully ionized. The number of free
electrons per helium nucleus ($I_{\rm He}$) is approximated as a simple
function of $T_{\rm eff}$. As a reasonable assumption, we adopt for OB
stars with $T_{\rm eff} > 20$~kK $I_{\rm He} = 2$, while for $T_{\rm
eff} < 20$~kK we adopt $I_{\rm He} = 1$ \citep{kudritzki89}. The
electron scattering opacity per unit mass (in units of
$\mathrm{cm^{2}g^{-1}}$) is then provided by
\begin{equation}
 \sigma_{\rm e} = 0.398 \, \frac{1 +  Y \,  I_{\rm He}}{1 + 4 \, Y }
,\end{equation}
where $Y$ is the surface helium number fraction, $Y = N_{\rm He}/N_{\rm H}$. 

Using the WLR, the mass-loss rates are then derived according to
\begin{equation}
 \log \, \dot{M} = \log \, D_{\rm mom} - \log \, v_{\infty} - \frac{1}{2} \, \log \, (R/R_{\odot}) 
\label{eq_wlr}
.\end{equation}
The parameters required to estimate both $D_{\rm mom}$ and
$v_{\infty}$ are obtained from the evolutionary calculations. In
addition, two further input parameters need to be provided, namely,
$\alpha'$ and $\log D_0$ (cf. Table~\ref{tab:xd}). The specification
of these parameters provides a simple way to calibrate the mass-loss
rates to observed WLRs, and to account for new observational or
theoretical results. 

Furthermore, it is convenient to apply a global scaling factor,  
denoted here as $f_{\rm scal}$, for the calculated
mass-loss rate, so that 
\begin{equation}
 \dot{M}_{\rm final} = f_{\rm scal} \cdot \dot{M}_{\rm calculated}.
\end{equation}
It is evident that similar results could be obtained with other
parameter settings, for example, by changing $\alpha'$ (cf.
Fig.~\ref{fig:alphas}) and/or $\log D_0$. However, while in the
following we mostly consider $f_{\rm scal}$ and $\alpha'$ as fixed, 
the specification of $\log D_0$ provides a simple way to account for
arbitrary changes in the mass-loss rates at the bi-stability jump.  

The implementation of the bi-stability jumps depends on the position
of the jump ($T_{\rm eff, jump}$), the method used to calculate the
mass loss around the jump (interpolation), and the size of the jump.
We emphasize that the observed behavior of the
ratio of $v_{\infty}/v_{\rm esc}$ over the jump is gradual
\citep{prinja98,cro2006} which implies that the change in $\dot{M}$
should be gradual as well \citep{markova2008}. 

\subsection{The position of the jump} 

Based on the discrepancy between
theoretical predictions from \cite{vink99} and observational results,
it is useful to control the position of the jump in terms of a jump
temperature $T_{\rm eff, jump}$. Theoretically, the jump temperature
is predicted to depend on the wind density, and \cite{vink2000}
calculate it via $\Gamma_{\rm e}$, while \cite{vink2001} calculate it
via $Z$. Since observations suggest that there may be a well-defined
$T_{\rm eff}$ where the bi-stability jump occurs (for a given
metallicity), we specify $T_{\rm eff, jump1}$ and $T_{\rm eff, jump2}$
as input parameters for the first and the second bi-stability jump
temperatures, respectively. This provides the flexibility to adjust these
parameters to observed or new theoretical values. Indeed, the very
recent study by \cite{petrov2016} indicates that the (theoretical)
jump temperature needs to be shifted toward lower effective
temperatures ($\approx 20 \, \mathrm{kK}$ in case of the first
bi-stability, in agreement with observations) than predicted
previously by \cite{vink99} and \cite{vink2000} ($\approx 25 \,
\mathrm{kK}$). This change will need to be adopted in future stellar
evolution models.

The bi-stability region itself is defined by its central
jump temperature ($T_{\rm eff, jump}$) and the half width of
interpolation ($\Delta T$). For simplicity, we have adopted the same
interpolation technique as present in the MESA Vink scheme. In
particular, a larger interpolation region will yield results similar
to a gradual change, while a small value of $\Delta T$ implies a steep
increase in $\dot{M}$. Considering observational constraints, we set
$T_{\rm eff, jump1} = 20,500 $ K, and use $\Delta T = 3,500 \,
\mathrm{K}$, unless otherwise stated. 
  
\subsection{The size of the jump} 

There are several parameters that
control the size of the jump. It is reasonable to consider that the
(input) parameter $f_{\rm vinf}$ decreases over the jump, based on the
observed ratio of terminal velocity and escape velocity
\citep{lamers95, prinja98, cro2006}. The adjustment of this parameter
directly influences the mass-loss rates. For example, if
$v_{\infty}/v_{\rm esc}$ steeply decreases by a factor of two
from the hot to the cool side of the jump (following the studies by
\citealt{lamers95,vink2000}), then, without further adjustment,
$\dot{M}$ would steeply increase by a factor of two as long
as the WLR is continuous. Since the change in $v_{\infty}/v_{\rm esc}$
is fairly well constrained by observations, and we intend to test the
effects of different behaviors of $\dot{M}$ alone, we use the
following parametrization for simplicity. 

We consider the offset value of the WLR at the hot side of the jump, $\log D_{0} (\rm hot)$, 
as fixed, and define a corresponding
\begin{equation}
\label{eq:wlr_d0}
  \log D_{0} (\rm cool) = \log D_{0} (\rm hot) + \Delta D_0
,\end{equation}
at the cool side of the jump, with $\Delta D_0$ an adjustable
parameter, allowing us to control the size of the jump in a simple and
flexible way.
 
\section{Results}

\subsection{The Vink mass-loss rates in evolutionary codes}
\label{sec:vinkevol}

\begin{figure}
\centering
  \resizebox{9cm}{!}{\includegraphics{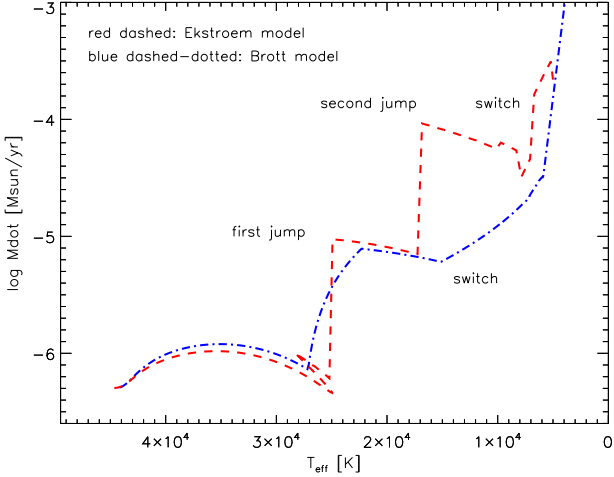}} 
\caption{Mass-loss histories for non-rotating Galactic 40 $\mathrm{M_{\odot}}$
models from \cite{ek12} and \cite{br11}, adopting the Vink mass-loss
prescription. See text.}
\label{fig:40_Mdots}
\end{figure}

In Fig.~\ref{fig:40_Mdots} we compare the implementation of the Vink
mass-loss recipe for the case of two non-rotating Galactic
40~$M_{\odot}$ models as computed by \cite{ek12} and \cite{br11}. This
plot shows one of the main foci of the present study: The
implementation of the first bi-stability jump predicts an increase in
the mass-loss rates by a factor of 15.4 in the \cite{ek12} model (at
25~kK) and by a factor of 10.7 in the \cite{br11} model (at
27--22~kK). The \cite{ek12} model results in a steep increase of
$\dot{M}$ across the first and the second bi-stability jumps. The
\cite{br11} model, on the other hand, uses a linear interpolation over
the first bi-stability jump region, noting that the expressions from
\cite{vink2001} do not account for the intermediate range between
22.5 and 27.5~kK. The latter method may provide a closer match to
observational constraints (Jorick Vink, priv.comm.), while the former
is not compatible with the behavior of mass-loss rates and terminal
velocities derived from observations (see Sect.~\ref{sec:muc}).

The second bi-stability jump is not implemented in the Bonn models. 
Instead, a switch is performed to the \cite{nie90} mass-loss rates
whenever $T_{\rm eff} < T_{\rm eff, jump1} \approx \, 25 \,
\mathrm{kK}$, and when the Vink rates would yield lower $\dot{M}$
values than the corresponding \cite{nie90} values (typically around
$16 \, \mathrm{kK}$). \cite{br11} argue that this strategy accounts
for the increased mass-loss rates at the second bi-stability jump. The
\cite{ek12} models, including the second bi-stability jump, apply the
\cite{vink2001} recipe until 12.5~kK, that is, to the minimum temperature
considered, and then switch to the \cite{dejager88} prescription (cf.
Fig.~\ref{fig:40_Mdots}). This yields mass-loss rates on the order of
$10^{-4} \mathrm{M_{\odot}}\mathrm{yr}^{-1}$ close to $T_{\rm eff} = 17$ kK, in
stark contrast with observations from typical B supergiants
\citep{cro2006,markova2008}. The values derived from observations are
typically two orders of magnitude lower than from these models. 

By estimating theoretical mass-loss rates based on a work-integral
method\footnote{Using radiative accelerations as calculated by the NLTE
atmosphere code CMFGEN \citep{Hillier98}}, \cite{petrov2016} found that
both the first and the second bi-stability jump should be located at
lower effective temperatures than predicted by \cite{vink2000}.
Namely, the first jump (Fe\,{\sc iv} recombining to Fe\,{\sc iii}) should
lie around 20~kK, and the second jump (Fe\,{\sc iii}
recombining to Fe\,{\sc ii}) around 9~kK. Note that these
values are much lower than the corresponding jump temperatures
in the evolutionary models displayed in Fig.~\ref{fig:40_Mdots}. 

\begin{figure}
\centering
  \resizebox{9cm}{!}{\includegraphics{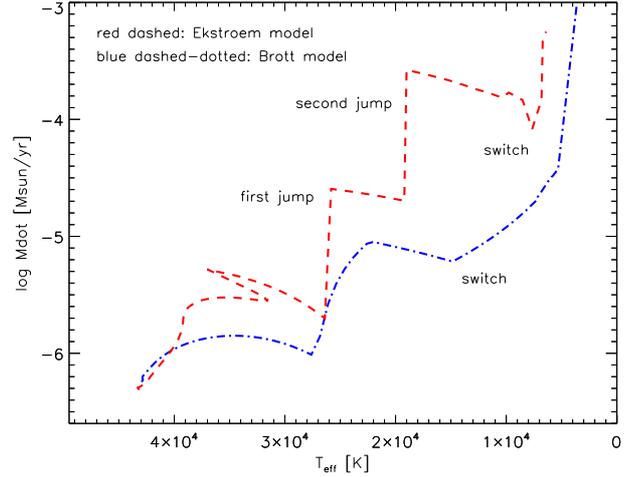}}  
\caption{As Fig.~\ref{fig:40_Mdots}, but for rotating models. Both
evolution models have $v_{\rm rot}({\rm initial})
\approx 315$ km s$^{-1}$. The Geneva model becomes more
luminous from $\approx$ 39~kK on, which results in higher mass-loss rates.}
\label{fig:40_Mdots_rot}
\end{figure}

The reason why a comparison of mass-loss rates from higher mass models
calculated by different numerical codes (using different assumptions
and parameters) is challenging is a consequence of the different
evolution in the HRD, which is more distinct for higher masses. The
main differences arise from overshooting, and in rotating models from
the treatment of angular momentum transport and chemical mixing. Since the
mass-loss rates have a strong dependence on luminosity, it is evident
that models with different luminosities will lead to different
mass-loss histories,
namely more luminous models will lose more mass. This effect can be
clearly seen in Fig.~\ref{fig:40_Mdots_rot} if one compares the
evolutionary tracks of the two models that include rotation, and
considers that the increase in luminosity of the 40 $\mathrm{M_{\odot}}$
\cite{ek12} track corresponds to an increase in the mass loss-rate
(beginning at around 39~kK). 

To conclude, with some dependence on the details of individual codes,
rotating models will show different mass-loss rates throughout their
evolution, even when adopting the same wind prescription. Referring to
the Vink rates, the size of the first bi-stability jump is barely
affected by the different mass-loss implementations though. 

MESA uses a mixture of the Bonn and Geneva approaches to implement the
Vink mass-loss rates. The first bi-stability jump is implemented in a
similar way to the Geneva code. A small difference concerns
the temperature region close to the jump. While the Geneva code uses
either Equation 14 or 15 from \cite{vink2001} to determine the
mass-loss rate directly at the hot and at the cool side, respectively,
the MESA implementation interpolates between the two equations, with
a very small half width of $\Delta T = 100 $~K.  

The second jump is discarded in the MESA implementation.
However, an alternative is incorporated in the code, allowing to
switch to other schemes at effective temperatures below the range of
applicability of the Vink rates ($<$12.5~kK). However, while the Bonn
group adopts such a switch whenever the Vink rates would result in
$\dot{M}$ lower than the \cite{nie90} rates, MESA conservatively
switches at 12.5~kK to any other mass-loss prescription specified by
the user. (We note that in the newest MESA release, r8118, this has been
changed, and the switch between the wind schemes can be set to occur
at a user-defined effective temperature.) This implies, considering
the properties of the Vink recipe, that the MESA implementation
results in a decreasing $\dot{M}$ from $\approx \, 16$ kK to 12.5 kK,
in contrast to the Bonn models that have an increasing mass loss in
this range. 


\subsection{The size of the first bi-stability jump in stellar evolution models}

The Vink formula predicts a large jump in the mass-loss rates from the
hot to the cool side of the jump, and this has often been quoted as an
average factor of 5 \citep{vink2000}. However, apart from a few
studies (e.g., \citealt{groh2014}), no further check of this statement
has been performed, namely whether evolutionary model
calculations actually result in such an average value. To this end, the
size of the jump as present in the Galactic \cite{ek12} and
\cite{br11} evolutionary grids is compared in Table \ref{tab:table3}.
For this comparison we have concentrated on non-rotating models, but
the results for rotating models are similar.

In a conservative approach, we considered the local minima on the hot
side, and the local maxima on the cool side, but using mean values on
both sides would result in similar ratios. The jump temperatures are
simply read off from the corresponding timestep, if the jump region is
not wider than 0.9~kK. This applies to almost all cases from
\cite{ek12}, except for the 20 and 25 $\mathrm{M_{\odot}}$ models at the second
bi-stability jump, and we suspect that the wider jump in these cases
is a result of a fast change in radius. The Bonn models apply a linear
interpolation for the transient $T_{\rm eff}$ regime, and thus the
corresponding jump temperature is provided as a range.

The outcome of our comparison is somewhat surprising. The average
increase in $\dot{M}$ over the first bi-stability jump corresponds to
a factor of 16.5 and 12.3 for the non-rotating Galactic \cite{ek12}
and \cite{br11} models, respectively. Almost all of the entries denote
an increase larger than a factor of 10. The 20~$\mathrm{M_{\odot}}$ models display
the largest jumps (a factor beyond 15), while the size of the jump
decreases for higher initial masses. For comparison, in the study by
\cite{vink99} a 20 $\mathrm{M_{\odot}}$ non-rotating Galactic model shows an
increase in $\dot{M}$ of only a factor of 6.5 (see their Figure
3). The source of this discrepancy is still unclear, but it is likely
that the actual stellar parameters at the position of the jump are
different from those adopted by \cite{vink99}. Nevertheless, these
values result in high $\dot{M}$ on the cool side of the jump, in stark
contrast with mass-loss rates derived from current
diagnostics (see previous sections). Therefore it seems that stellar
evolution models might significantly overestimate the mass-loss rates
after the bi-stability jump(s).  

One further difficulty relates to overshooting. The Geneva models with
smaller step overshooting always reach the TAMS before the first
bi-stability jump, while the Bonn models reach the TAMS at lower
effective temperatures, because of the larger overshoot parameter
(see, e.g., their published HRDs). However, thus far it is unclear
what happens when the end of the main sequence and the first
bi-stability jump occur simultaneously. We need to  understand
whether there might be a physical interaction between the significant
internal changes and the mass loss driven by the wind. Stellar parameters do change rapidly upon reaching the TAMS, and
the mass-loss rates will change accordingly. If the TAMS coincides
with the jump temperature a more complex behavior may occur. This
might also become important if one additionally accounts for the
accompanying angular momentum-loss (bi-stability braking, see
Sect.~\ref{sec:bsb}).
\begin{table*}
\caption{Comparison of the increase of mass-loss rates over the
bi-stability jump(s) in the non-rotating Galactic \cite{ek12} and
\cite{br11} evolutionary grids. }
\centering
\begin{tabular}{c|c c c c|c c}
\hline\hline
& \multicolumn{4}{ c }{Geneva} & \multicolumn{2}{ |c }{Bonn} \\
\hline
 M [$\mathrm{M_{\odot}}$] & $T_{\rm jump1}$ [kK] & $\frac{\dot{M}_{\rm
 cool}}{\dot{M}_{\rm hot}}$ (1)& $T_{\rm jump2}$ [kK] &
 $\frac{\dot{M}_{\rm cool}}{\dot{M}_{\rm hot}}$ (2)& $T_{\rm jump1}$
 [kK] & $\frac{\dot{M}_{\rm cool}}{\dot{M}_{\rm hot}}$ (1) \\
 
20 & 23.5 & 19.6 & 15.7 - 12.9 & 11.9 & 26.0 - 20.5 & 16.6 \\

25 & 24.0 & 17.9 & 14.6 - 10.4 & 10.2 & 26.3 - 21.5 & 15.7 \\

30 & - & - & - & - & 26.6 - 21.7 & 12.1 \\

32 & 24.3 & 16.6 & 15.7 & 13.3 & - & - \\

35 & - & - & - & - & 26.9 - 21.8 & 11.2 \\

40 & 24.9 & 15.4 & 16.8 & 13.1 & 27.2 - 22.3 & 10.7 \\

50 & 25.1 & 14.0 & 17.8 & 13.2 & 27.6 - 22.7 & 10.0 \\

60 & 25.4 & 13.1 & 18.1 & 12.6 & 27.9 - 23.1 & 9.6 \\

average &  & 16.5 &  & 12.2 &  & 12.3 \\
\hline
\end{tabular}
\tablefoot{Ratios of $\dot{M}$ on the cool 
and hot side (i.e., below and above the jump temperature) exceed
a factor of 10. Rotating models show similar ratios.}
\label{tab:table3}
\end{table*}

\subsection{Pre-bi-stability behavior (PBB)}

\begin{figure} 
\centering
  \resizebox{9cm}{!}{\includegraphics{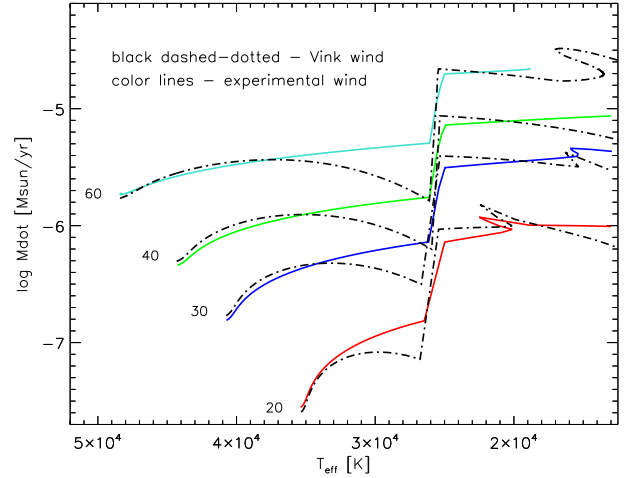}}  
\caption{Comparison between MESA models applying either the Vink recipe or
our experimental wind scheme, in terms of mass-loss rates versus
effective temperature. Initial masses (in solar units) indicated
next to the tracks. See text.
\label{fig:fig3}}
\end{figure}

In our first step, we aimed at calibrating our experimental wind routine
to recover the Vink rates, with particular emphasis on the first
bi-stability jump. Our description (see end of Sect.~\ref{sec:muc})
allows for an approximate reproduction of the jumps (as a function of
$M_{\rm init}$) as computed by the Vink wind scheme
(Fig.~\ref{fig:fig3}). For the complete mass range considered, 20 - 60
$\mathrm{M_{\odot}}$, good agreement (at least when concentrating on the
average behavior) was obtained when using $f_{\rm vinf} =
v_{\infty}/v_{\rm esc} = 2.6$ on the hot side of the first jump,
$f_{\rm vinf} = 1.3$ on the cool side, and simultaneously increasing
the WLR offset at the cool side by $\Delta D_0 = 0.35$ (see
Eq.~\ref{eq:wlr_d0}). This choice of parameters corresponds to an
average increase in $\dot M$ by a factor of $\approx 4.5$. Additional
parameters for this test are as follows. We adopted $x = 1.83$
(following the theoretical value provided by \citealt{vink2000}),
corresponding to $\alpha' \approx 0.55$.  The jump temperature was
fixed at $T_{\rm eff, jump1} = 25 \, \mathrm{kK}$, and the second
bi-stability jump had been ignored. All models were calculated for
Galactic metallicity and without rotation.
At the hot side of the jump and towards higher temperatures, however, 
an important qualitative difference in the behavior of the mass-loss
rates needs to be discussed. While the mass-loss rates from our
experimental routine monotonically increase from higher $T_{\rm eff}$
until the jump, the Vink rates display a curvature, with a local
maximum well before the jump temperature (see Fig.~\ref{fig:fig3}).
This difference relates to the fact that the Vink recipe has a strong
dependence on $T_{\rm eff}$, being the potentially largest factor
influencing $\dot M$ on the hot side of the jump, $\log \dot M \propto
10.92\, \cdot \, \{\log (T_{\rm eff}/40000 )\}^2$. Thus, the Vink
prescription is calibrated at 40~kK, which roughly corresponds to the
ZAMS temperature of a 30 $\mathrm{M_{\odot}}$ Galactic star. This scaling keeps
the temperature dependence small around temperatures close to 40~kK,
while close to the bi-stability jump this dependence dominates and decreases $\dot{M}$ compared to higher values of $T_{\rm
eff}$.
Furthermore, since $\dot{M}$ decreases before the bi-stability jump,
the size of the jump is effectively larger than if such a decrease was
not present. 

On the other hand, our wind tool includes only an indirect dependence
on $T_{\rm eff}$, via $L$ and $R$. As long as there is no explicit
dependence via defining $\log D_0 = \log D_0(T_{\rm eff})$ (and we
refrained from including such a dependence in the present study),
$\dot M$ mainly depends on luminosity\footnote{when combining the
second and third term on the rhs of Eq.~\ref{eq_wlr} and assuming
$v_\infty \propto v_{\rm esc}$, any explicit radius dependence
vanishes}, and because of the monotonic increase of this quantity for
our 40~$\mathrm{M_{\odot}}$ model (see Fig.~\ref{fig:comp_hrd_g}), the mass-loss
rate also increases towards the jump.

Thus, the most important qualitative difference between the Vink rates and
our experimental wind scheme is their ``pre-bi-stability behavior'' (PBB).
We define the PBB as the behavior of the mass-loss rates at effective
temperatures higher than the first bi-stability jump temperature, and the
starting point of the PBB is where the mass-loss rates derived from the Vink
formula start to decrease with decreasing $T_{\rm eff}$ (and thus where the
Vink and the experimental wind begin to depart qualitatively). This is, of
course, initial-mass dependent: for larger masses, the PBB will start
earlier, at higher $T_{\rm eff}$ (see Figure \ref{fig:fig3}). 

Although one can easily identify the source of this difference,
originating from the specific temperature scaling of the Vink
prescription, a corresponding decrease in the mass-loss rate with decreasing
$T_{\rm eff}$ cannot be identified from the semi-empirical WLR 
as used here (i.e., with constant $D_0$), due to the dominating
effect of increasing luminosity (see above).

Even though the luminosity increases, \cite{vink2000} argue that there
is a physical explanation why the line acceleration should become less
effective at lower $T_{\rm eff}$ (when considering the range between
50 and 30~kK): Due to the shift of the flux maximum towards longer
wavelengths, the number of effective lines decreases. With respect to
our approach, this would mean that $D_0$ should decrease as well.
Unfortunately, there is no strict observational evidence to support
either scenario \citep{cro2006,markova2008,fraser2010}. This means that it is not
established whether the mass-loss rates increase or decrease with
$T_{\rm eff}$ in the PBB region. In order to discriminate between the two
cases, a meaningful analysis of mass-loss rates (and not only
wind-momenta) should be performed in this quite narrow $T_{\rm eff}$
range, for a significant sample of massive stars. 

In any case, a comparison between WLRs derived from observations and model
calculations is a non-trivial task. While from observations a sample
of stars with different (initial and actual) masses can be analyzed at
a certain point in their evolution, evolutionary models provide the
complete path of a stellar model for a given initial mass. In
particular, any WLR derived from a grid of evolutionary models with
different masses will diminish the actual PBB seen when concentrating
on individual tracks. 
Such an effect can be already noted in the wind momentum vs.
luminosity diagram presented by \citet[their Figure 9, upper
panel]{vink2000}, where individual patterns (e.g., the $\dot{M}(T_{\rm
eff})$ behavior) are smeared out, and an almost strictly linear relation over
the complete range between 50 and 27.5~kK ``survives'', consistent
with our approach of adopting a constant $D_0$. 

When interpreting theoretical or observed WLRs, potential degeneracies
(same luminosity, but other parameters different) need to be taken
into account as well. For instance, HD\,210809 and HD\,15629 are two
O stars with similar luminosities \citep{rep2004}, and the derived
$\dot{M}$ is higher for HD\,210809, which is the cooler object. This
would be consistent with model calculations if the cooler star was
less massive and more evolved. For instance, our 30 and $40 \,
\mathrm{M_{\odot}}$ models with the experimental wind scheme may be appropriate
for such a case. These two models reach the same luminosity at 
$T_{\rm eff} = 27.5 \, \mathrm{kK}$ and $T_{\rm eff} = 44 \,
\mathrm{kK}$, respectively. At these positions, the $\dot{M}$ of the
cooler object is slightly larger indeed, when predicted following our
approach (see Fig.~\ref{fig:fig3}).

\begin{table}
\caption{Increase of mass-loss rates over the first
bi-stability jump in non-rotating Galactic models
using the Vink and the experimental wind scheme.}
\centering
\begin{tabular}{ccc}
\hline\hline
 & Vink wind & experimental wind \\
 \hline 
 M [$\mathrm{M_{\odot}}$] & $\dot{M}_{\rm cool} / \dot{M}_{\rm hot}$ &  $\dot{M}_{\rm cool} / \dot{M}_{\rm hot}$ \\

20 & 14.5 & 5.0   \\

30 & 14.0 & 4.5 \\

40 & 13.8 & 4.1 \\

60 & 13.4 & 3.9 \\
\hline
\end{tabular}
\label{tab:bsj}
\end{table} 

The PBB has a large impact on the size of the
jump in mass-loss rate, when considering the immediate region in 
$T_{\rm eff}$ enclosing the jump temperature. To provide a numerical
comparison between the Vink and the experimental wind scenario,
Table~\ref{tab:bsj} displays the increase of $\dot{M}$ from the hot to
the cool side of the jump. For the considered masses between 20 to 60
$\mathrm{M_{\odot}}$, the average increase of $\dot{M}$ is a factor of 13.5
using the Vink recipe, while it is 4.4 for the experimental wind
(consistent with the adopted change in $v_{\infty}/v_{\rm esc}$ by a
factor of two, and $\Delta D_0 = 0.35$~dex; note that for this test
the experimental wind rates have been calibrated to match the Vink
rates at early phases and at the cool side of the bi-stability jump).
In other words, the increase of $\dot{M}$ during the PBB results in a
much smaller jump at the bi-stability than predicted by the Vink
recipe when considering individual tracks.

We add here that within our prescription of the experimental wind,
the choice of a sufficiently large $\Delta T$ (increasing the width of
the jump) would lead to similar effects as produced by the
\cite{br11} interpolation of mass-loss rates between 27.5 and 22.5~kK
(see above). In particular, such a procedure would
also diminish the pre-bi-stability decrease in $\dot{M}$ in the Vink
recipe, leading to a somewhat smaller effective jump (cf. 
Figs.~\ref{fig:40_Mdots} and \ref{fig:40_Mdots_rot}).
It might be useful to consider such a choice of $\Delta T$ for future studies.

\subsection{The experimental wind scheme: impact of $\alpha'$}

\begin{figure} 
\centering
\resizebox{9cm}{!}{\includegraphics{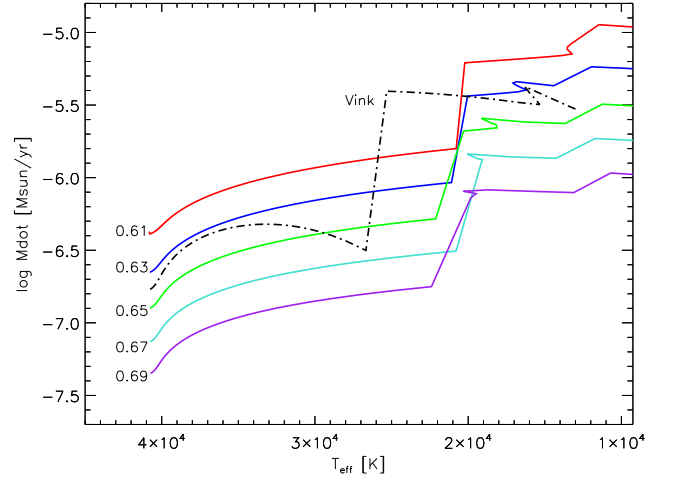}}   
\caption{Mass-loss histories for non-rotating Galactic 30
$\mathrm{M_{\odot}}$ MESA models, adopting different values for the
WLR-parameter related to its slope, $\alpha'$ = 0.61 to 0.69
(indicated next to the tracks). A model with the same initial
parameters but using the Vink recipe (without the second jump) is
shown for reference. 
\label{fig:alphas}}
\end{figure}

There are four global parameters ($\alpha'$, $\log\, D_0$, $f_{\rm vinf}$,
and $f_{\rm scal}$) in our setup that can influence the calculated
mass-loss rates, and in Fig.~\ref{fig:alphas} we show an important
test case for a 30 $\mathrm{M_{\odot}}$, Galactic $Z$, non-rotating model, when
$\alpha'$ is varied (we recall that this parameter has a physical
meaning, and is most commonly adopted in the range between 0.50 and
0.70), and the other parameters remain unchanged. Following the
clumping-corrected values from \cite{mokiem2005} and
\cite{mokiem2007a}, $\alpha'$ has been specified to lie in the range
between 0.61 and 0.69 (corresponding to $x = 1.63 - 1.45$), while
$\log D_0$ has been fixed at a value corresponding to smooth winds, 
$\log D_0 = 18.40$. Obviously, the choice of $\alpha'$ has a
significant impact on the produced mass-loss rates: even for the
rather small range of $\alpha'$ considered, $\dot M$ varies by a
factor of $\ga$10. Similar results would have been obtained if
$\alpha'$ was fixed, and $\log D_0$ was varied. This degeneracy of
$\dot M$ with respect to $\alpha'$ and $D_0$ is an important issue; for example, when
adopting values which are both corrected for wind-inhomogeneities, the
mass-loss rates might become significantly reduced compared to the
displayed situation (see Sect.~\ref{sec:intro} and below). 

To obtain an impression of the overall mass-loss history, we follow
the evolution until the coolest regime of line-driven winds. In these
models, the first bi-stability jump temperature has been set to
$T_{\rm eff, jump1} = 20.5$~kK, according to observational constraints,
and we adopted $\Delta D_{0} (1) = 0.35$, following our calibration
from the previous section. Interestingly, a variation of $\alpha'$
affects also the position of the TAMS, because of different mass-loss
rates: when lowering $\alpha'$ and thus increasing $\dot M$, the TAMS
is shifted to lower effective temperatures.

Moreover, we also considered the second jump, at a temperature $T_{\rm
eff, jump2} = 12$~kK. For example, when assuming a continuous WLR for late
B and early A-type supergiants (i.e., $\Delta D_{0} (2)= 0$), and a decrease
in $v_{\infty}/v_{\rm esc}$ from 1.3 to 0.7 (e.g.,
\citealt{markova2008}), this second jump would produce only a minor
increase of the mass-loss rates. 

\subsection{The experimental wind scheme: jump properties}
 
\begin{figure} 
\centering
\resizebox{9cm}{!}{\includegraphics{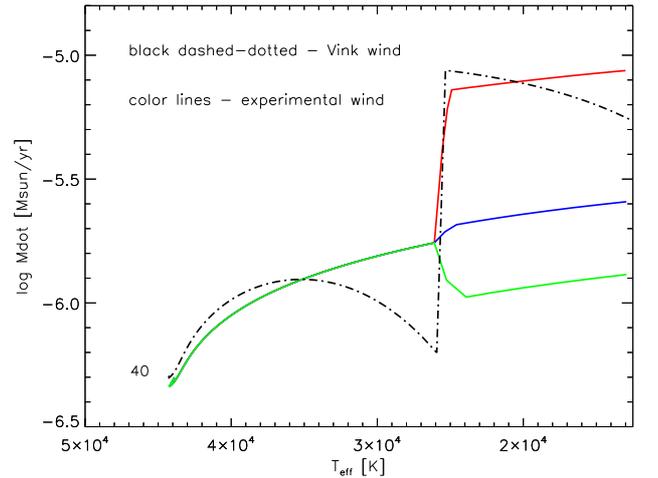}}  
\caption{Mass-loss histories for non-rotating Galactic 40
$\mathrm{M_{\odot}}$ MESA models, with different bi-stability jump properties.
Red: increase in $\dot M$ ($\Delta D_0 = 0.35$); blue: (almost)
continuous $\dot M$ ($\Delta D_0 = - 0.20$); green: moderate decrease
in $\dot M$ ($\Delta D_0 = - 0.50$).  A model with the same initial
parameters but using the Vink recipe (without the second jump) is
shown for reference.}
\label{fig:jump}
\end{figure}

To test the response to different $\dot{M}$ prescriptions at the first
bi-stability jump, we again considered non-rotating Galactic models.
Figure~\ref{fig:jump} shows a corresponding 40 $\mathrm{M_{\odot}}$ MESA model,
where alternative bi-stability scenarios have been simulated, by
varying the offset of the WLR over the jump, quantified by $\Delta
D_0$. Here, we used again a ``high'' $T_{\rm eff, jump1} = 25 \,
\mathrm{kK}$, and we simulated increasing, continuous, and moderately
decreasing mass-loss rates, by adopting $\Delta D_0 =~0.35, -0.20$,
and $-0.50$, respectively.  The latter two scenarios are in agreement
with the results from \cite{markova2008}, while a significant jump
corresponds to the \cite{vink99} predictions. We note that for a
continuous $\dot M$ over the bi-stability region, the offset of the
WLR must be decreased to compensate for the decrease in
$v_{\infty}/v_{\rm esc}$,
\begin{equation}
 \dot{M}({\rm hot}) \, \propto \frac{D_{\rm mom}({\rm hot})}
 {v_{\infty}/v_{\rm esc}({\rm hot})} \, =: \, \dot{M}({\rm cool}) \, \propto 
 \frac{D_{\rm mom}({\rm cool})}{v_{\infty}/v_{\rm esc}({\rm cool})}. 
\end{equation}

\subsection{Reducing the mass-loss rates in rotating models}
\label{subsec:comp_rot}
 
After the simple parameter tests presented above, we turn our
attention to rotating Galactic models. In the following model
calculations, we set the initial rotational velocities to $v_{\rm
rot}({\rm initial}) = 300 \, \mathrm{kms^{-1}}$, and we varied only
the overall scaling factor ($f_{\rm scal} = 1.0, 0.6, 0.3$), while
keeping $\alpha' = 0.543$ and $\log D_0 = 18.40$, respectively. For
the mass range (20-60 $\mathrm{M_{\odot}}$) and metallicity ($Z = 0.014$)
considered in this work, this setup recovers the Vink rates when
$f_{\rm scal} = 1$.

\begin{figure} 
\centering
\resizebox{9cm}{!}{\includegraphics{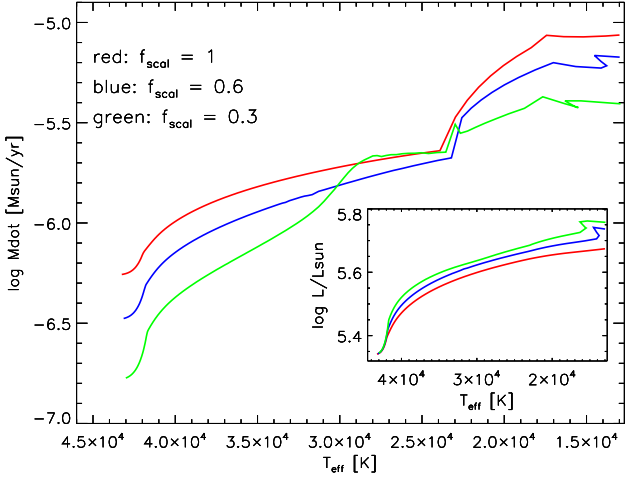}}  
\caption{Mass-loss histories for rotating Galactic 40 $\mathrm{M_{\odot}}$
MESA models, with $v_{\rm rot}$(initial) = 300~km s$^{-1}$, and
different mass-loss rates corresponding to scaling factors $f_{\rm
scal} = 1.0$ (red), $f_{\rm scal} = 0.6$ (blue), and $f_{\rm scal} =
0.3$ (green). The corresponding HRD is provided as an inset.
See text for further details.} 
\label{fig:rotdot}
\end{figure}

Fig.~\ref{fig:rotdot} shows the result of this test, which aims at
simulating the effects from globally reduced mass-loss rates (compared to
presently used values), a scenario which is quite likely in view of
the current debate (cf. Sect.~\ref{sec:intro}). 

It is immediately clear that the inclusion of rotation has a very
large impact on the resulting mass-loss rates and vice versa. This
effect, however, is rather complex. Close to the ZAMS, the mass-loss
rates indeed correspond to their reduced values, and in non-rotating models these differences would remain preserved
throughout the evolution, as shown, for example, in our $\alpha'$ parameter
study (Fig.~\ref{fig:alphas}). In the rotating models, on the other
hand, the initially different mass-loss rates converge to almost
identical values between 28-23~kK, before diverging again at lower
temperatures. Since we have used quite a large value of $\Delta T =
3500 \, \mathrm{K}$ when interpolating over the bi-stability region,
the corresponding change in $\dot M$ is rather gradual (again, we
adopted $\Delta D_0 (1)= 0.35$ to recover the Vink mass-loss rate at
the cool side of the jump).

\begin{figure} 
\centering
\resizebox{9cm}{!}{\includegraphics{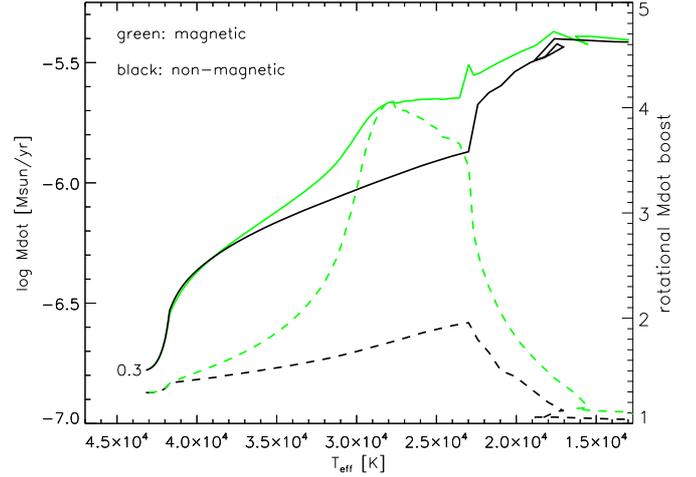}}  
\caption{As Fig.~\ref{fig:rotdot}, for two models with $f_{\rm scal} =
0.3$. The only difference between the models is the inclusion of
magnetic-field effects from an internal Spruit-Tayler dynamo on the
angular momentum transport (green line; this model is
identical to the $f_{\rm scal} = 0.3$ model from
Fig.~\ref{fig:rotdot}), and the exclusion of such effects
(black line). Additionally, we show the corresponding
rotational $\dot{M}$ boost factors (dashed lines and right ordinate).}
\label{fig:rotmag}
\end{figure}

To investigate the reason for this interesting behavior, most 
prominent in the model with $f_{\rm scal} = 0.3$, we turned off the
Spruit-Tayler dynamo which otherwise helps to preserve the surface
angular momentum due to a coupled core-envelope configuration.
Figure~\ref{fig:rotmag} displays the outcome of this test, namely that the internal effects related to rotation can, indeed, have an influence on
the (reduced) mass-loss rates. 

The large difference between the mass-loss rates from the coupled and
uncoupled models originates, in fact, from their rotational
properties. When the Spruit-Tayler (ST) dynamo is used for angular
momentum transport (as in \citealt{br11}), the model behaves as a
solid body rotator (this process also induces an efficient
chemical mixing, because of an efficient Eddington-Sweet
circulation), therefore it must maintain a flat internal angular
velocity profile (see \citealt{paxton2013}, their figure 29).  When
this mechanism is not considered, the core angular velocity is higher
than in the envelope. 

The reason that the originally reduced mass-loss rates become higher
than anticipated (the model with $f_{\rm scal} = 0.3$
reaches mass-loss rates as high as the model with $f_{\rm scal} = 1$, see Fig.~\ref{fig:rotdot}), is the ``rotational $\dot{M}$ boost'' 
\citep{paxton2013} implemented in MESA (see Eq.~\ref{eq:rotf_b}), 
which approximately accounts for an increased mass-loss due to 
centrifugal acceleration. We remind the reader that the \cite{br11}
models account for a similar mechanism. Additionally, we
note that due to these differences the two models do not evolve at
identical luminosities, though the corresponding differences are
very small ($\le$ 0.02~dex until 15~kK).

\begin{figure} 
\centering
\resizebox{9cm}{!}{\includegraphics{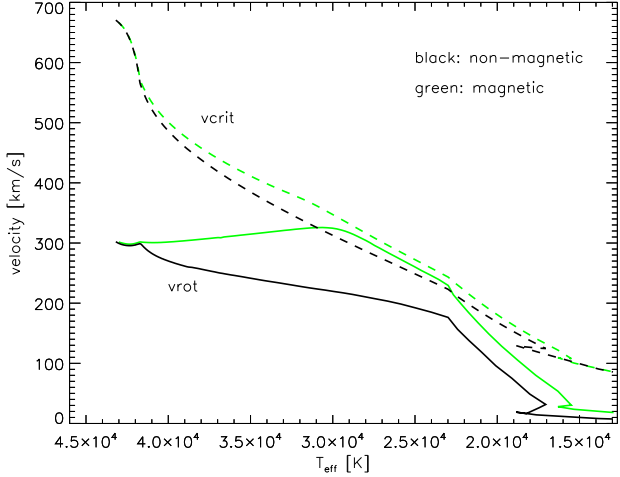}}  
\caption{Rotational and critical velocities for the two models from 
Fig.~\ref{fig:rotmag}.}
\label{fig:rotvel}
\end{figure}

The boost itself results from a significant difference in the
development of the surface rotational velocities of the ST coupled
(magnetic) and uncoupled (non-magnetic) models. From
Fig.~\ref{fig:rotvel}, we see that the quasi solid body rotator (i.e.,
the ST coupled model, green line) approaches its critical velocity
around 30~kK. While the rotational velocity remains almost
constant, the critical velocity decreases, mostly because of an
increasing radius (and, in the problematic formulation of
Eq.~\ref{eq:vcrit_b}, also because of the increase in luminosity).
The reason for the almost constant surface rotational velocity
in the ST coupled model is that the surface angular momentum lost due
to the wind can be efficiently replaced by angular momentum extracted
from the core, thus keeping the surface rotation high. In other words,
when strong coupling is present, the whole star must be braked, not
only the surface.

When the rotational velocity approaches the critical velocity, the approximate expression for
the rotational boost of $\dot M$ becomes inadequate, but we refrained
from manipulating the formulation adopted in MESA -- this would
require a separate investigation of its own. 

Nevertheless, since the basic effects should be qualitatively correct,
we conclude that in these test cases an original reduction (referring
to slow rotation) of the mass loss, and hence angular momentum loss,
is able to significantly influence stellar evolution. In this regard,
the treatment of internal angular momentum transport (strongly
affected by the presence or absence of internal magnetic fields) 
plays a major role. Therefore, the revision of mass-loss rates (from the perspective of angular momentum loss)
cannot be studied separately, since when changing the angular momentum
transport, significant feedback effects will influence the angular momentum
loss.

As a side note, we remark that the model with $f_{\rm scal} = 0.3$ and
internal magnetic field displays only a marginal bi-stability jump in
$\dot M$. In this case, the rotational boost is more efficient at the
hot than at the cool side, since at the cool side the rotational
velocity departs from the critical one, presumably due to the
significant mass loss and angular momentum loss just before and across the
jump. Thus, the change in mass loss over the jump is weaker than if
both sides would be similarly affected. Further studies on this issue
might be required, since it might partly explain a jump lower than
predicted, assuming that indeed the mass-loss rates (when
discarding rotational effects) were weaker than presently adopted. 

\subsection{The evolution of surface rotational velocities}
\label{subsec:evol_vrot}

If the mass-loss rates are reduced, the angular momentum loss
decreases under typical conditions, which has a severe effect on the rotational properties at all stages
of stellar evolution. This is one of the issues that can be
constrained via observations of $v \sin i$, particularly for main
sequence and blue supergiant stars. Later and pre-supernova phases are
affected even more, since the actual mass and angular momentum content
determines the final fate of the star. 

\begin{figure} 
\centering
 \resizebox{9cm}{!}{\includegraphics{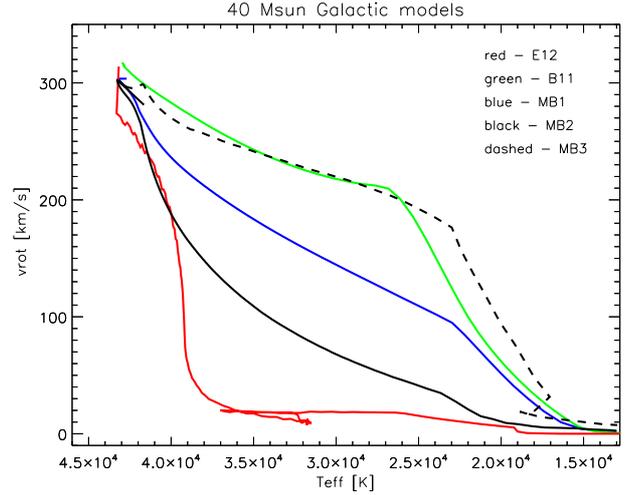}}
\caption{Surface rotational velocities vs. effective temperature, for
rotating Galactic models at $40 \mathrm{M_{\odot}}$. Models shown are published by \cite{br11} (B11, green line), \cite{ek12} (E12, red
line), and three MESA models (MB1, blue line; MB2, black line; MB3,
black dashed line). See text.}
\label{fig:fig5}
\end{figure}

To investigate the evolution of $v_{\rm rot}$ in various
scenarios, we compare, in a step-by-step approach, our MESA models
utilizing the experimental wind scheme with the \cite{ek12} and
\cite{br11} tracks (Fig.~\ref{fig:fig5}, red and green lines,
respectively). 

First, we display a $40 \, \mathrm{M_{\odot}}$ ($Z = 0.014$) model (MB1, blue
line) that has similar characteristics as the corresponding one from
\cite{br11}. In particular, we adopted (i) $\alpha_{ov} = 0.335$, (ii)
Spruit-Tayler dynamo generated magnetic fields when accounting for the
angular momentum transport, (iii) meridional circulation in a
diffusive approach, and (iv) mixing efficiencies $f_c = 0.0228$ and
$f_{\mu} = 0.1$. The (experimental) mass-loss rates have been
calibrated in such a way that they roughly agree, in earlier phases
and at the cool side of the first bi-stability jump, with the Vink
rates, though the PBB and consequently the change of $\dot M$ over the
jump are different. 

A departure between the rotational velocities as predicted by our MESA
and the \cite{br11} model is already seen at early phases (around
$40 \, \mathrm{kK}$). We speculate that these early differences are a
consequence of differences in the implementation accounting for the
effects of a dynamo generated field in the radiative zones
\citep{heger2005, petrovic2005}. We note, for example, that without magnetic
fields the rotational velocity decreases even more drastically in
early phases, as obvious from the corresponding \cite{ek12} model.
The differences in the slope from $35 \, \mathrm{kK}$ on might be
associated with the different PBB, that is, the differences in $\dot{M}
(T_{\rm eff})$, of the Vink rates and our experimental wind scheme (see
Fig.~\ref{fig:fig3}). Since the Bonn models have a shorter MS lifetime (compared to
analogous MESA models), they evolve faster, and lose less mass (thus
less angular momentum). This issue may explain why their $v_{\rm rot}$
values are consequently higher on the MS compared to our MESA models.
Further differences relate to a cooler bi-stability jump temperature
in the experimental wind scheme (see next section). 

It is well known that internal magnetic fields are predicted to have a major
impact on the evolution of surface rotational velocities (see, e.g.,
Fig.~3 from \citealt{maeder2005}, and also \citealt{maeder2009}).
Based on our previous finding that internal magnetic fields can also
affect mass-loss rates, via angular momentum transport, we calculated
a second model without internal magnetic fields (MB2, black line).
Evidently, this model displays considerably lower $v_{\rm rot}$ during its
complete MS evolution, compared to the magnetic one. At early stages,
this model has a similar $v_{\rm rot}$ history as the corresponding
model from \cite{ek12}.

Subsequently, we decreased the overall mass-loss rates in the latter
non-magnetic model to 30\% of its original value (MB3, black dashed
line). Very interestingly, this modification results in a similar
evolution of $v_{\rm rot}$ as in the original Bonn model. (This
similarity could, of course, be also achieved by a somewhat weaker
reduction of the mass-loss rates in the magnetic model). 
The large effect of the reduced mass-loss rates on the surface
rotational velocities is remarkable. Though \cite{maeder2005} pointed
out that the loss of angular momentum at the surface has a limited
impact on the internal rotational properties, here we have
shown that the surface angular momentum loss has a large impact on the
observable rotational velocities, though it also depends on the
effects of internal transport mechanisms (coupled vs. uncoupled
configuration). Thus, the evolution of the surface rotational
velocities can only be studied if both the surface angular momentum
loss and the internal angular momentum transport are considered in a
realistic manner. Conversely, a study of the surface rotational
velocities can provide severe constraints on these issues.

One may now ask to what extent these findings depend on initial mass. To
this end, we calculated a small grid of models with similar
assumptions as above, for the range of 20 - 60 $\mathrm{M_{\odot}}$. In the
following section, we quantify the strong impact of $\dot{M}$ on
$v_{\rm rot}$ as a function of initial mass, and discuss the outcome
of our simulations in the context of bi-stability braking proposed by
\cite{vink2010}. 

\subsection{The need for bi-stability braking}
\label{sec:bsb}

For Galactic O-type stars, the measured (projected) rotational
velocities display a large scatter, reaching up to $v_{\rm rot}~\sin i~
\approx~400$~km~s$^{-1}$ (e.g., \citealt{howarth97}),  though
the average initial rotational velocities of massive O stars are
currently debated. For example, \citealt{simon2014} find that essentially 
all O supergiants of their northern Galactic sample and 71\% of all O
dwarfs therein have $v_{\rm rot} \sin i <~200$~km~s$^{-1}$.
In the \cite{howarth97} sample, a significant drop in the
rotational velocities of B supergiants is observed at around 22~kK. To date, there is overwhelming observational evidence supporting an
average surface rotational velocity on the order of $v_{\rm rot} \sin
i \approx 50$ km s$^{-1}$ for late blue supergiants
\citep{huang2010,hunter2009,fraser2010}. If the blue supergiants are
the direct descendants of O-type stars, then the steep drop in the rotational velocities implies that a braking mechanism must be
present. To this end, \cite{vink2010} proposed that a large jump in
the mass-loss rates at the first bi-stability (the bi-stability
braking, BSB) could efficiently remove surface angular momentum, and
hence reduce the surface rotation of the star. However, we stress
again that at least a large jump in $\dot{M}$ is debated, and
that the jump temperature which is used in current evolutionary models
most likely needs to be shifted to cooler temperatures.

Moreover, \cite{vink2010} emphasized that the possibility of
BSB depends on the adopted evolutionary models. However, it has not
been investigated in detail which models would require a BSB to reach
low $v_{\rm rot}$ in the B supergiant regime, and which would not.

\begin{figure*}
\centering
\includegraphics[width=0.48\textwidth]{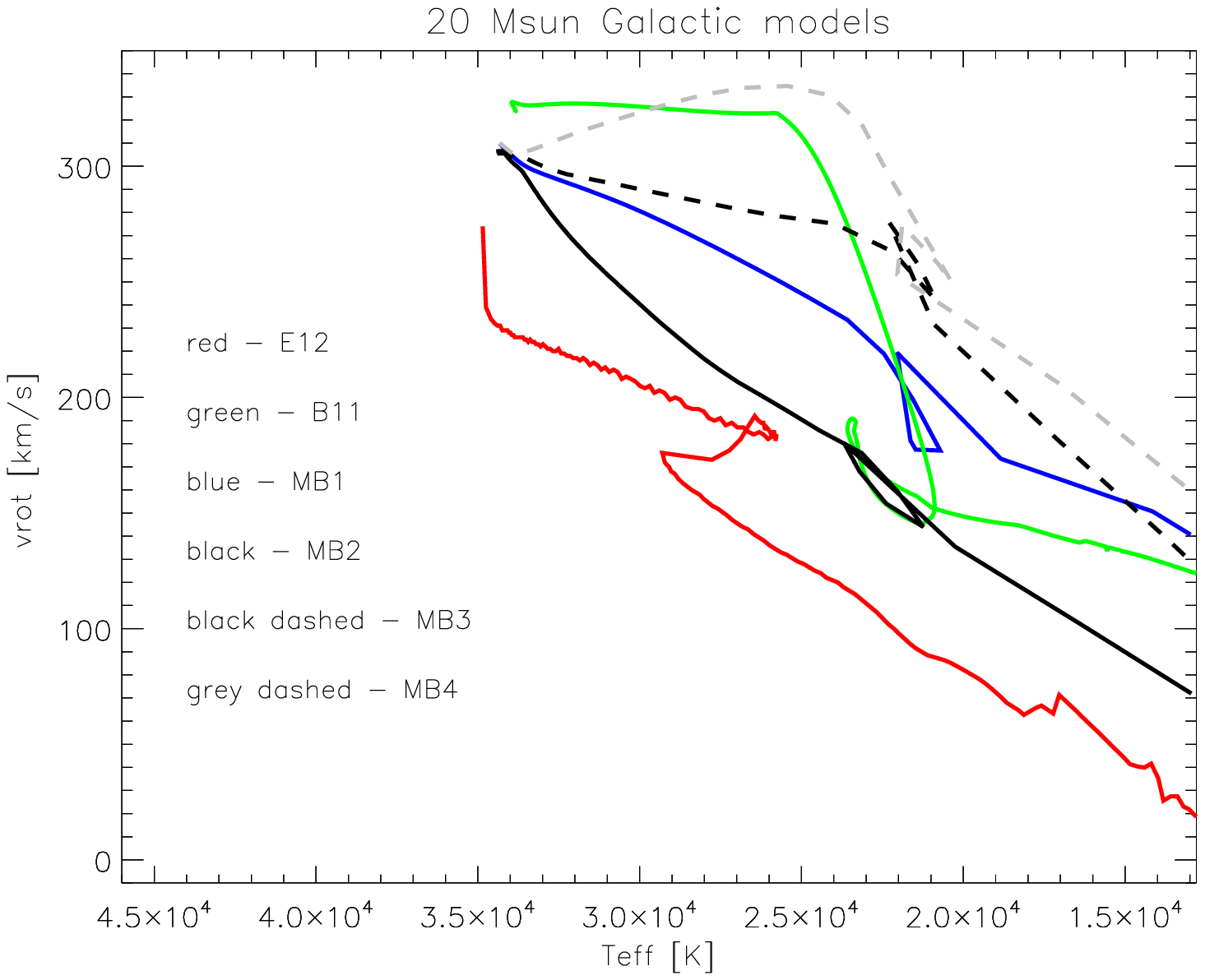}\hspace{0.03\textwidth}\includegraphics[width=0.48\textwidth]{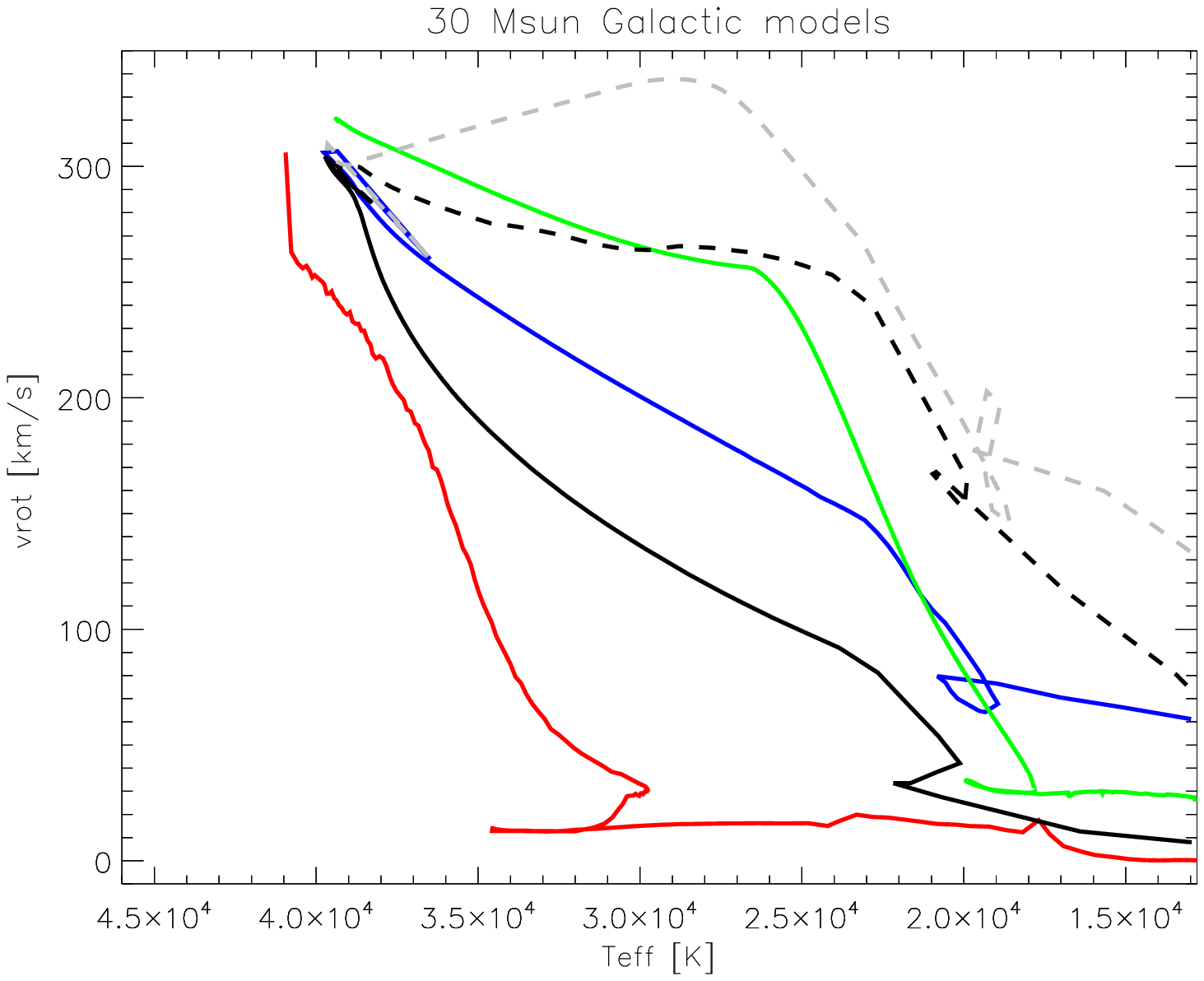}
\vspace{0.25cm}
\includegraphics[width=0.48\textwidth]{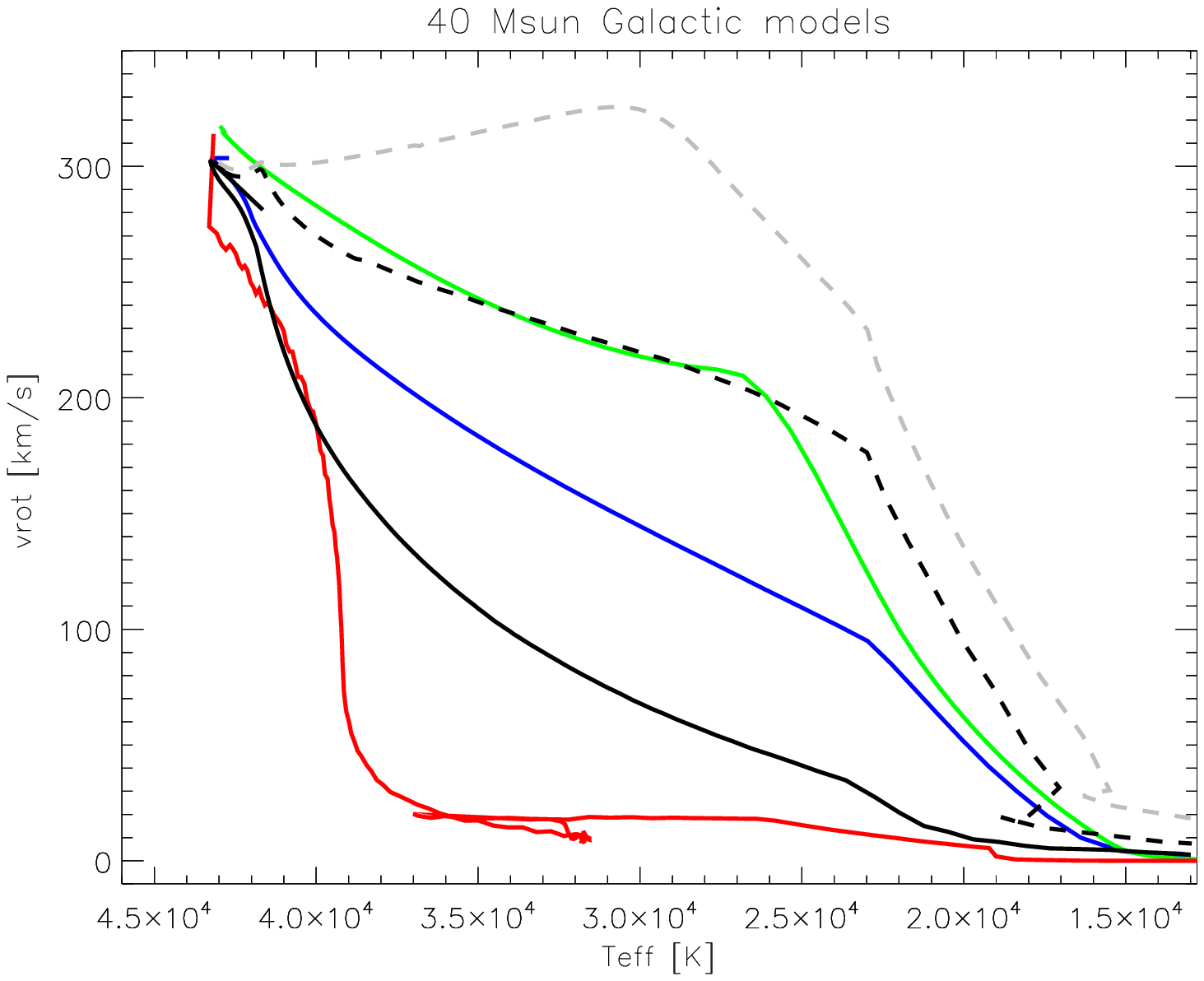}\hspace{0.03\textwidth}\includegraphics[width=0.48\textwidth]{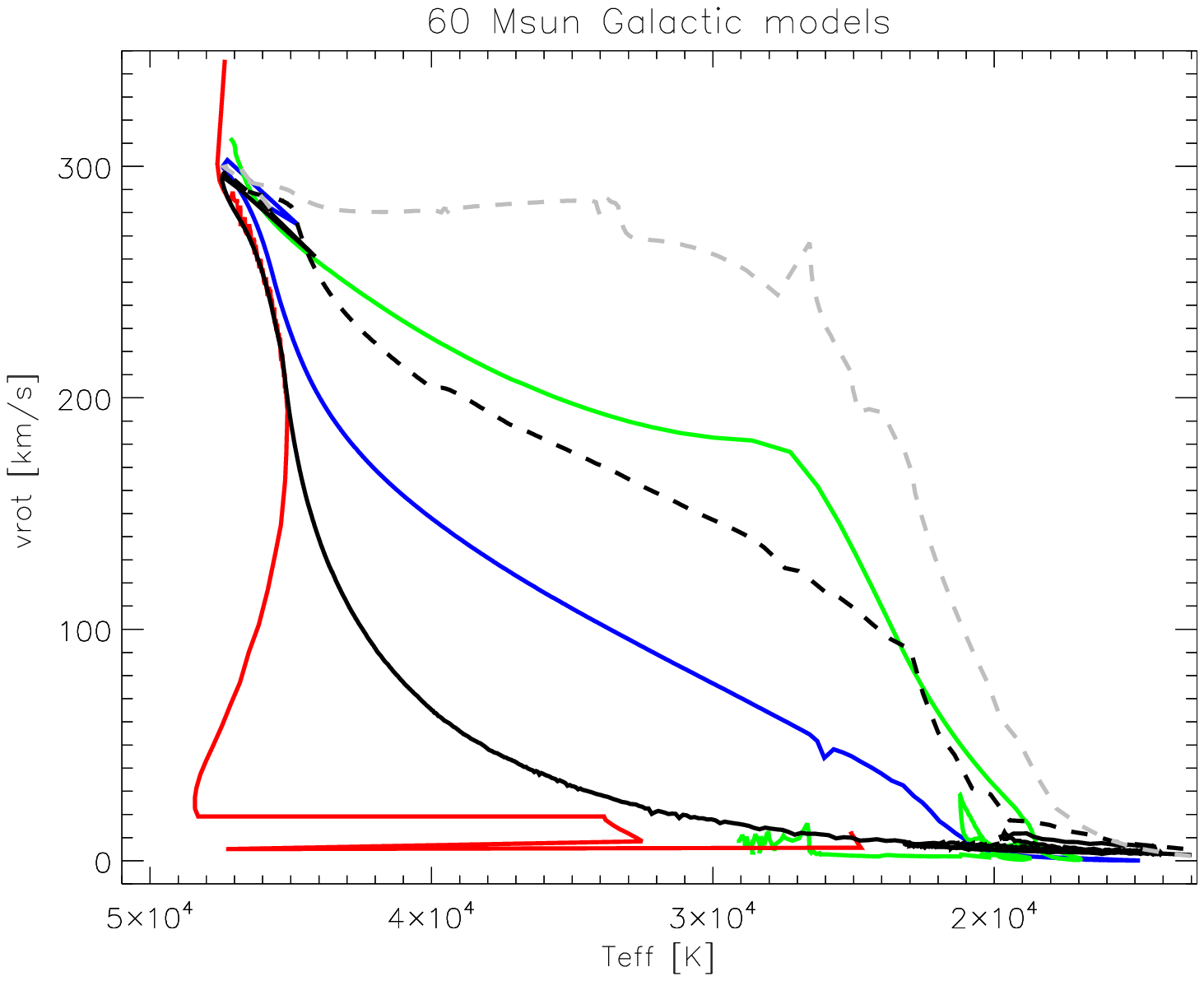}
\caption{Various Galactic metallicity models to investigate the
potential role of bi-stability braking, with initial masses between 20
to 60 $M_\odot$, and initial rotational velocities $\approx 300$ km s$^{-1}$. For reference, the Geneva models (red, E12) and
the Bonn models (green B11) are indicated. (Note that there is no 30 $\mathrm{M_{\odot}}$ but a
32 $\mathrm{M_{\odot}}$ model available from the \cite{ek12} grid.)
For the set-up of MESA models MB1 to MB4, see Table~\ref{tab:bsb1} and
Sect.~\ref{subsec:evol_vrot}.}
\label{fig:fig6}
\end{figure*}

The 32 - 60 $\mathrm{M_{\odot}}$ tracks of \cite{ek12} can be easily inspected
(Fig.~\ref{fig:fig6}). These models do not need a BSB; they already
brake the surface rotational velocities at high temperatures, because
of discarding internal magnetic fields and applying Vink mass-loss
rates, which in most cases are larger than those used in the Bonn
models, due to a higher luminosity (see Sect.~\ref{sec:vinkevol}). The
20 and 25 $\mathrm{M_{\odot}}$ tracks (the latter not shown here) are somewhat
misleading: Because of the comparatively small overshoot parameter,
these tracks reach the end of the main sequence just before the
bi-stability jump. Here, we interpret the hooks in the $v_{\rm
rot}$-tracks as a result from the corresponding hooks in the HRD,
meaning those related to changes in the internal structure of the star, and
not as a result of bi-stability braking. This assumption can be
justified if one recalls that the adopted jump temperature is at 24
and 25~kK for the 20 and 25 $\mathrm{M_{\odot}}$ model, respectively. These
values clearly do not coincide with the drop in $v_{\rm rot}$ that
occurs at higher temperatures, therefore the braking in this case is
attributed to reaching the TAMS. We conclude that none of the
\citet{ek12} models would require a BSB to account for slowly rotating
B supergiants.  One might now argue that these models, with weak
overshooting, would yield lifetimes in the B supergiant regime that
are too short to be compatible with the observed large population of
such objects, at least when relying on the hypothesis that they were
the immediate descendants of O stars. However, even when increasing
the overshoot parameter to account for this possibility, the large
angular momentum loss in early phases would not be affected, and still
no BSB would be required. 

The \cite{br11} models maintain a higher surface angular
momentum before the first jump, mainly due to the coupled
core-envelope configuration achieved by the effects of a Spruit-Tayler
dynamo mechanism in the radiative zone (as discussed before in
Sect.~\ref{subsec:comp_rot}). For most of these models, it can be safely
stated that bi-stability braking is required to reduce the
rotational velocities and to match the observed values. Only the 20
$\mathrm{M_{\odot}}$ case is problematic. Even with an increase of a factor of
15 in the mass-loss rates, the bi-stability jump seems to be
insufficient to brake this model's large surface rotational velocity. 
To investigate  the impact of important assumptions regarding
internal momentum transport and mass loss (which obviously cannot be
tested when relying on published models alone), we calculated a small
grid of MESA models with different setup, labeled as MB1 to MB4,
respectively. The general setup of models MB1 to MB3 has already been described in the
previous section, and these models are augmented here by an additional
magnetic model MB4 with $f_{\rm scal} = 0.3$. These models are summarized in Table~\ref{tab:bsb1} for easy comparison.

Evaluating the results displayed in Fig.~\ref{fig:fig6}, the role of
the BSB can be determined. As a conclusion on its necessity, the following
(crude) criterion was checked for each model: is there a steep drop in
the surface rotational velocity required to obtain models in the
region $T_{\rm eff} < 20$ kK and $v_{\rm rot} < 100$ kms$^{-1}$? In
other words, would the models evolve to $T_{\rm eff} < 20$ kK with
$v_{\rm rot} > 100 \, \mathrm{kms^{-1}}$ without the benefit of an
increase in $\dot{M}$ related to the bi-stability of the winds? If the
answer to both questions is yes, then the BSB may be required - keeping
in mind that the BSB is not the only mechanism capable of reducing
angular momentum. By inspecting Fig.~\ref{fig:fig6} (see also the last column of
Table \ref{tab:bsb1}), we find the following situation: whereas in
models MB1 and MB2 a BSB is only required for stars with $M < 30 \,
\mathrm{M_{\odot}}$, models with a decreased mass-loss rate (MB3 and MB4,
respectively) would always require a BSB - except for the 60 $\mathrm{M_{\odot}}$
MB3 model -- to enable slowly-rotating B supergiants.  
\begin{table}
\caption{Various models to check the requirement for BSB.}
\centering
\begin{tabular}{c|c|c|c}
\hline\hline
model & $f_{\rm scal}$  & internal B field & BSB required for \\
\hline
MB1 & 1 & yes &  $<$ 30 $\mathrm{M_{\odot}}$ \\
MB2 & 1 & no &  $<$ 30 $\mathrm{M_{\odot}}$\\
MB3 & 0.3 & no &  $<$ 60 $\mathrm{M_{\odot}}$ \\
MB4 & 0.3 & yes &  all models\\
\hline
\end{tabular}
\tablefoot{ $f_{\rm scal} = 1 $ refers to our experimental wind scheme
with parameters that approximately recover the Vink rates at early
phases and at the cool side of the first bi-stability jump (see
Sect.~\ref{subsec:evol_vrot}). Magnetic fields are accounted for via
the Spruit-Tayler dynamo as implemented in MESA.}
\label{tab:bsb1}
\end{table} 
However, there is also the problem that in almost all 20
$\mathrm{M_{\odot}}$ models (except for the one presented by \citealt{ek12}),
and in models MB3/MB4 with 30 $\mathrm{M_{\odot}}$, the bi-stability braking
alone is still not sufficient to push the rotational velocities below
$100 \, \mathrm{km/s}$ for $T_{\rm eff} < 20$~kK. This is a serious
challenge, as we note that the increase in $\dot{M}$ as adopted here may
already overestimate the actual situation. 

\section{Discussion}

As shown by our model calculations, the evolution of the surface rotational velocities of massive stars are not
only determined by the effects of internal angular momentum transport,
but are strongly and qualitatively influenced by the magnitude and evolution of
mass-loss rates. Although the existence and origin of internal magnetic fields, and their role in
determining the internal angular momentum transport are debated, such
magnetic fields - when included in evolutionary models - often dominate this transport mechanism  
\citep{maeder2005, heger2005}.
Moreover, the actual treatment of the Eddington-Sweet circulation
(i.e., a diffusive or advective approach) must also be considered when
relying on specific evolutionary models, since in a diffusive approach
this is the largest term contributing to chemical mixing (see also
\citealt{song2016}). 

Regardless of the differences in the treatment of meridional
circulation, mixing efficiencies, or the consideration of internal
magnetic fields, a reduction of the mass-loss rates (i.e. $f_{\rm
scal} = 0.3$) compared to the currently used theoretical prescriptions
($f_{\rm scal} = 1$) would result in less angular momentum loss, which
in turn would keep the surface rotation in massive stars closer to
their initial velocities throughout the main sequence
evolution. According to our simulations, only for $M \ga 60 \, \mathrm{M_{\odot}}$ it is still possible to brake stellar rotational velocities
significantly, when internal magnetic fields are not included (see
Fig.~\ref{fig:fig6}). This is a
very drastic and observable effect, especially as the
mass-loss rates were changed by a factor of only between two and three when
accounting for the different overall rates.
  
In our models with a fully diffusive scheme, the current
disagreement between mass-loss rates predicted by the standard Vink recipe and those derived from observations leads to a dichotomy
of potential scenarios:
\begin{itemize}
\item[(i)]If the mass-loss rates derived from
observations are considered (a factor of between two and three less than the Vink rates),
then there is a need for bi-stability braking, requiring a
large jump in $\dot{M}$ which has not been observed thus far.
Otherwise, model calculations would predict a large sample of late B
supergiants with high surface rotational velocities. 
\item[(ii)] If, on the other hand, the Vink rates 
(for $T_{\rm eff} > T_{\rm eff, jump1}$) are correct, there may be no
need for a large jump in $\dot{M}$ at the bi-stability (depending on
the particular evolutionary model), in agreement with observational
evidence. 
\end{itemize}

In this context, we must also consider (at least) two observational
constraints. Namely, is the average surface rotational velocity
adopted at the ZAMS (300 km s$^{-1}$) too high? Or, is it possible that
there are late B supergiants with $v_{\rm rot} > 100$ km s$^{-1}$ that are not observed, for example, due to their short
lifetime in that region? 

Indeed, it is not clear that all O stars are fast rotators at or close to
the ZAMS. As already pointed out, the majority of northern Galactic O
stars analyzed by \cite{simon2014} has $v_{\rm rot} \sin i < 200$ km s$^{-1}$. If these values were common, then for a
considerable number of stars, mass-loss rates lower (by a factor of
between two and three) than those currently adopted would mean that BSB is not required.
In this case, no significant angular momentum loss would be required
to brake rotation and thus to enable the production of slowly rotating
B supergiants. 

Whether a larger number of late B supergiants with $v_{\rm rot} > 100$ km s$^{-1}$ exists but has not been detected might also be
an observational issue. Adopting a large overshoot parameter for
consistency reasons in our model calculations from \cite{br11} ensures
that our late B supergiants are still core hydrogen burning, although
the typical main sequence lifetime of the models below 20~kK is short.
For example, in case of our 40 $\mathrm{M_{\odot}}$ models (Fig.~\ref{fig:fig6}) only
the last 3\% of their main sequence lifetime is spent in that regime. 

There is no clear consensus whether the observed B supergiants are 
core-hydrogen burning main sequence or core-helium burning post-main
sequence objects (e.g., \citealt{vink2010,meynet2015}). Indeed, it
might be possible that B supergiants are not a homogeneous sample,
which would make the evaluation of a requirement for BSB even more
challenging. Nevertheless, if B supergiants are the direct
descendants of O stars, a physical mechanism for angular momentum loss
must be established, whether or not the BSB exists.

\section{Conclusions and future work}

In this study we have investigated the impact of mass loss on the early
stages of massive star evolution. We aimed to understand whether
the discrepancy between mass-loss rates from theoretical predictions
and from recent diagnostics could be clarified in terms of
evolutionary constraints. To this extent, we developed a simple wind
routine which has been implemented into MESA, and we simulated
stellar evolution using various mass-loss rates, particularly rates
that are either compatible with those predicted by \citet{vink2000} or
with state-of-the-art observational diagnostics.

Our experimental wind description is based on the semi-empirical
WLR, and has been implemented within an
easily adjustable, flexible and fast routine. For the sake
of simplicity, we considered the corresponding parameters, slope $x$ =
$1/\alpha'$ (see Equation \ref{eq:alf}), and offset $\log D_0$, as
constant for $T_{\rm eff} > T_{\rm eff, jump1}$. For more
sophisticated models a $T_{\rm eff}$-dependence might need to be
considered (e.g., \citealt{lamers95, puls2000}). Furthermore, due to
the line-driving mechanism, both parameters depend on metallicity,
though in this work we only considered Galactic models, thus avoiding
such an explicit dependence. We emphasize that this observationally
guided routine relies on existing scaling relations, and that it is
not a new mass-loss description, in contrast to the much more complex
approaches by, for example, \cite{graefener2005}, \cite{bouret2012}, or
\cite{petrov2016}.

From our model calculations, in which we adopted a fully diffusive scheme, we draw the following conclusions that are mostly independent of assumptions about internal magnetic fields and their coupling with angular momentum transport: For a mass range between 20 to 60 $\mathrm{M_{\odot}}$, and canonical initial rotational speeds, $v_{\rm rot}({\rm init})
\approx$ 300~km s$^{-1}$, it is not possible to simultaneously
account for (i) lower overall mass-loss rates (here: by a factor of between two and three),
and (ii) a smaller increase of $\dot{M}$ over the bi-stability region,
compared to the predictions by \cite{vink2000, vink2001}. Otherwise,
the models would retain too high surface rotational velocities in the
late B supergiant regime. An obvious alternative would become feasible
if either the rotational velocities at/close to the ZAMS were
significantly lower than adopted here, or if a yet unidentified,
efficient braking mechanism would operate during the early stages of
massive star evolution. 

As an interesting secondary result, we also found that initially weaker
winds can become significantly amplified in their subsequent
evolution by the rotational boost, for models which account for a
coupled core-envelope configuration due to magnetic fields (quasi
solid body rotators). This effect would lead to a lower effective
$\dot M$ jump across the bi-stability region, and might help to
understand corresponding observational findings.

During our investigation, we identified the following problems that
must be studied to enable further progress: 
\begin{itemize}
\item What are the real mass-loss rates before the onset of the 
      bi-stability, for temperatures lower than roughly 35~kK? Do they
      increase (experimental wind) or decrease (Vink rates) with time?
      This pre-bi-stability behavior (PBB) plays a crucial role in
      determining the actual value of the bi-stability jump regarding
      $\dot M$.  
\item Is there a gradual change in mass-loss rates at 
      $\approx 18 - 23 \, \mathrm{kK}$, corresponding to the observed
      gradual behavior of $v_{\rm esc}/v_{\infty}$?
\item What are the average ZAMS surface rotational velocities of 
      O-type stars, and are there any rapidly-rotating (single) B
      supergiants below $20 \, \mathrm{kK}$? 
\item Approximately 10\% of massive stars have detected surface
      magnetic fields (e.g., \citealt{wade2014}), and thus could
      experience magnetic braking accounting for slow rotation (see \citealt{meynet2011}). Is it
      possible that weaker, as-yet undetected surface magnetic fields in
      massive stars result in an efficient removal of surface angular
      momentum? The challenge of detecting magnetic fields in the often complex
      and variable spectra of O stars makes this question worthy of further
      investigation.
\item Stellar evolution codes do not agree on the implementation 
      of the first and second bi-stability jump
      adopting the Vink rates, and the size of the jump in $\dot{M}$
      is much larger than the originally considered value by
      \cite{vink2000}. This is in contradiction with observations (and
      also with recent wind calculations by \citealt{petrov2016}), and
      evolutionary models might significantly overestimate the
      mass-loss rates at the cool side of the bi-stability.
\item The necessarily simplified treatment of the complex interactions 
      and evolution of magnetic fields in 1D hydrodynamical
      calculations implies that results will be approximate. While the
      characteristics of internal fields and their effects are still
      debated, it is clear that such fields, if present, could have
      extremely important consequences for stellar evolution.
      Moreover, the existence of strong surface magnetic fields is now
      know for a handful of O stars (e.g., \citealt{wade2015}). The
      direct effects of these surface magnetic fields - on both mass
      loss and rotation - also urgently need to be implemented in
      stellar evolution calculations (e.g., Petit et al., in prep).
\end{itemize}

Although, in our opinion, the predicted evolution vs. diagnostics of
$v_{\rm rot}$ is an ideal tool to study open issues in evolutionary
calculations, particularly regarding mass loss, there are also other
constraints that need to be considered. Most important in this respect
are diagnostics of the abundances of nuclear-processed material mixed into
the stellar surface layers, where the mixing efficiency strongly depends on
the angular velocity profile between the core and the envelope. Since
diagnostics of surface CNO abundances have made considerable
progress in recent years (e.g., \citealt{hunter2008, przybilla2010,
rivero2012, bouret2012, martins2015a, martins2015b}), the inclusion of
such results into studies similar to the present work will certainly
lead to further understanding of the evolution of massive stars during
the main sequence and beyond.

\begin{acknowledgements}
We appreciate fruitful discussions with Jorick Vink, Ines Brott,
Norbert Langer, Raphael Hirschi, and Jose Groh, and we acknowledge the
MESA developers for making their code publicly available. We
thank the referee of this paper, Cyril Georgy, for his valuable
comments and suggestions. GAW acknowledges Discovery Grant support
from the Natural Science and Engineering Research Council (NSERC) of
Canada.
\end{acknowledgements}

\bibliography{keszthelyi_et_al}

\begin{thebibliography}{122}
\expandafter\ifx\csname natexlab\endcsname\relax\def\natexlab#1{#1}\fi

\bibitem[{Abbott(1982)}]{abbott82}
Abbott, D.~C. 1982, The Astrophysical Journal, 259, 282

\bibitem[{Aerts(2015)}]{aerts2015}
Aerts, C. 2015, in New windows on massive stars: asteroseismology,
  interferometry, and spectropolarimetry, Vol. 307, Proceedings of the
  International Astronomical Union, IAU Symposium, 154--164

\bibitem[{Angulo {et~al.}(1999)Angulo, Arnould, Rayet, Descouvemont, Baye,
  Leclercq-Willain, Coc, Barhoumi, Aguer, Rolfs, Kunz, Hammer, Mayer,
  Paradellis, Kossionides, Chronidou, Spyrou, degl Innocenti, Fiorentini,
  Ricci, Zavatarelli, Providencia, Wolters, Soares, Grama, Rahighi, Shotter, \&
  Lamehi~Rachti}]{angulo99}
Angulo, C., Arnould, M., Rayet, M., {et~al.} 1999, Nuclear Physics A, 656, 3

\bibitem[{Arnett {et~al.}(2009)Arnett, Meakin, \& Young}]{arnett2009}
Arnett, D., Meakin, C., \& Young, P.~A. 2009, The Astrophysical Journal, 690,
  1715

\bibitem[{Asplund {et~al.}(2005)Asplund, Grevesse, \& Sauval}]{asplund2005}
Asplund, M., Grevesse, N., \& Sauval, A.~J. 2005, in Astronomical Society of
  the Pacific Conference Series, Vol. 336, Cosmic Abundances as Records of
  Stellar Evolution and Nucleosynthesis, ed.: T.G Barnes III and F.N. Bash, 25

\bibitem[{{Bjorkman} \& {Cassinelli}(1993)}]{Bjorkman93}
{Bjorkman}, J.~E. \& {Cassinelli}, J.~P. 1993, \apj, 409, 429

\bibitem[{Bouret {et~al.}(2012)Bouret, Hillier, Lanz, \&
  Fullerton}]{bouret2012}
Bouret, J.-C., Hillier, D.~J., Lanz, T., \& Fullerton, A.~W. 2012, Astronomy \&
  Astrophysics, 544, 67

\bibitem[{Braithwaite(2006)}]{braithwaite2006}
Braithwaite, J. 2006, Astronomy \& Astrophysics, 449, 451

\bibitem[{Briquet {et~al.}(2007)Briquet, Morel, Thoul, Scuflaire, Miglio,
  Montalban, Dupret, \& Aerts}]{briquet2007}
Briquet, M., Morel, T., Thoul, A., {et~al.} 2007, MNRAS, 381, 1482

\bibitem[{Briquet {et~al.}(2012)Briquet, Neiner, Aerts, Morel, Mathis, Reese,
  Lehmann, Costero, Echevarria, Handler, Kambe, Hirata, Masuda, Wright, Yang,
  Pintado, Mkrtichian, Lee, Han, Bruch, De~Cat, Uytterhoeven, Lefever,
  Vanautgaerden, de~Batz, Fr{\'e}mat, Henrichs, Geers, Martayan, Hubert, Thizy,
  \& Tijani}]{briquet2012}
Briquet, M., Neiner, C., Aerts, C., {et~al.} 2012, MNRAS, 427, 483

\bibitem[{Brott {et~al.}(2011)Brott, de~Mink, Cantiello, Langer, de~Koter,
  Evans, Hunter, Trundle, \& Vink}]{br11}
Brott, I., de~Mink, S.~E., Cantiello, M., {et~al.} 2011, Astronomy \&
  Astrophysics, 530, 115

\bibitem[{Castor {et~al.}(1975)Castor, Abbott, \& Klein}]{cak75}
Castor, J.~I., Abbott, D.~C., \& Klein, R.~I. 1975, Astronomy \& Astrophysics,
  195, 154

\bibitem[{Castro {et~al.}(2014)Castro, Fossati, Langer, Sim{\'o}n-D{\'i}az,
  Schneider, \& Izzard}]{castro2014}
Castro, N., Fossati, L., Langer, N., {et~al.} 2014, Astronomy \& Astrophysics,
  570, 13

\bibitem[{Chieffi \& Limongi(2013)}]{chieffi2013}
Chieffi, A. \& Limongi, M. 2013, The Astrophysical Journal, 764, 21

\bibitem[{Cohen {et~al.}(2013)Cohen, Sundqvist, \& Leutenegger}]{cohen2013}
Cohen, D., Sundqvist, J., \& Leutenegger, M. 2013, Massive Stars: From alpha to
  Omega, held 10-14 June 2013 in Rhodes, Greece, 1, 36

\bibitem[{Crowther {et~al.}(2006)Crowther, Lennon, \& Walborn}]{cro2006}
Crowther, P.~A., Lennon, D.~J., \& Walborn, N.~R. 2006, Astronomy \&
  Astrophysics, 446, 279

\bibitem[{Cunha {et~al.}(2006)Cunha, Hubeny, \& Lanz}]{cunha2006}
Cunha, K., Hubeny, I., \& Lanz, T. 2006, The Astrophysical Journal, 647, 143

\bibitem[{Cyburt {et~al.}(2010)Cyburt, Amthor, Ferguson, Meisel, Smith, Warren,
  Heger, Hoffman, Rauscher, Sakharuk, Schatz, Thielemann, \&
  Wiescher}]{cyburt2010}
Cyburt, R.~H., Amthor, A.~M., Ferguson, R., {et~al.} 2010, The Astrophysical
  Journal Supplement Series, 189, 240

\bibitem[{de~Jager {et~al.}(1988)de~Jager, Nieuwenhuijzen, \& van~der
  Hucht}]{dejager88}
de~Jager, C., Nieuwenhuijzen, H., \& van~der Hucht, K.~A. 1988, Astronomy \&
  Astrophysics Supplement Series, 72, 259

\bibitem[{de~Mink {et~al.}(2009)de~Mink, Cantiello, Langer, Pols, Brott, \&
  Yoon}]{demink2009}
de~Mink, S.~E., Cantiello, M., Langer, N., {et~al.} 2009, Astronomy \&
  Astrophysics, 497, 243

\bibitem[{Eggenberger {et~al.}(2008)Eggenberger, Meynet, Maeder, Hirschi,
  Charbonnel, Talon, \& Ekstr{\"o}m}]{eggenberger2008}
Eggenberger, P., Meynet, G., Maeder, A., {et~al.} 2008, Astrophysics and Space
  Science, 316, 43

\bibitem[{Ekstr{\"o}m {et~al.}(2012)Ekstr{\"o}m, Georgy, Eggenberger, Meynet,
  Mowlavi, Wyttenbach, Granada, Decressin, Hirschi, Frischknecht, Charbonnel,
  \& Maeder}]{ek12}
Ekstr{\"o}m, S., Georgy, C., Eggenberger, P., {et~al.} 2012, Astronomy \&
  Astrophysics, 537, 146

\bibitem[{{Espinosa Lara} \& {Rieutord}(2011)}]{espinosa2011}
{Espinosa Lara}, F. \& {Rieutord}, M. 2011, \aap, 533, A43

\bibitem[{{Feldmeier} {et~al.}(2003){Feldmeier}, {Oskinova}, \&
  {Hamann}}]{Feldmeier03}
{Feldmeier}, A., {Oskinova}, L., \& {Hamann}, W.-R. 2003, \aap, 403, 217

\bibitem[{Fraser {et~al.}(2010)Fraser, Dufton, Hunter, \& Ryans}]{fraser2010}
Fraser, M., Dufton, P.~L., Hunter, I., \& Ryans, R.~S.~I. 2010, Monthly Notices
  of the Royal Astronomical Society, 404, 1306

\bibitem[{Friend \& Abbott(1986)}]{friend86}
Friend, D.~B. \& Abbott, D.~C. 1986, The Astrophysical Journal, 311, 701

\bibitem[{Georgy {et~al.}(2013)Georgy, Ekstr{\"o}m, Granada, Meynet, Mowlavi,
  Eggenberger, \& Maeder}]{georgy2013}
Georgy, C., Ekstr{\"o}m, S., Granada, A., {et~al.} 2013, Astronomy \&
  Astrophysics, 553, 24

\bibitem[{{Gr{\"a}fener} \& {Hamann}(2005)}]{graefener2005}
{Gr{\"a}fener}, G. \& {Hamann}, W.-R. 2005, \aap, 432, 633

\bibitem[{Grevesse {et~al.}(1996)Grevesse, Noels, \& Sauval}]{grevesse96}
Grevesse, N., Noels, A., \& Sauval, A.~J. 1996, in Astronomical Society of the
  Pacific Conference Series, held in: Astrophysics Conference in College Park;
  Maryland; 9-11 October 1995; San Francisco., Vol.~99, Proceedings of the
  sixth annual October Astrophysics Conference, ed.: by S.S. Holt and G.
  Sonneborn, 117

\bibitem[{Groenewegen {et~al.}(1989)Groenewegen, Lamers, \& Pauldrach}]{groe89}
Groenewegen, M.~A.~T., Lamers, H.~J.~G.~L.~M., \& Pauldrach, A.~W.~A. 1989,
  Astronomy \& Astrophysics, 221, 78

\bibitem[{Groh {et~al.}(2014)Groh, Meynet, Ekstr{\"o}m, \& Georgy}]{groh2014}
Groh, J.~H., Meynet, G., Ekstr{\"o}m, S., \& Georgy, C. 2014, Astronomy \&
  Astrophysics, 564, 30

\bibitem[{Heger {et~al.}(2000)Heger, Langer, \& Woosley}]{heger2000}
Heger, A., Langer, N., \& Woosley, S.~E. 2000, The Astrophysical Journal, 528,
  368

\bibitem[{Heger {et~al.}(2005)Heger, Woosley, \& Spruit}]{heger2005}
Heger, A., Woosley, S.~E., \& Spruit, H.~C. 2005, The Astrophysical Journal,
  626, 350

\bibitem[{Herv{\'e} {et~al.}(2013)Herv{\'e}, Rauw, \& Naze}]{herve2013}
Herv{\'e}, A., Rauw, G., \& Naze, Y. 2013, Astronomy \& Astrophysics, 551, 83

\bibitem[{Herwig(2000)}]{herwig2000}
Herwig, F. 2000, Astronomy \& Astrophysics, 360, 952

\bibitem[{{Hillier}(1991)}]{Hillier91}
{Hillier}, D.~J. 1991, \aap, 247, 455

\bibitem[{{Hillier} \& {Miller}(1998)}]{Hillier98}
{Hillier}, D.~J. \& {Miller}, D.~L. 1998, \apj, 496, 407

\bibitem[{Hirschi {et~al.}(2004)Hirschi, Meynet, \& Maeder}]{hirschi2004}
Hirschi, R., Meynet, G., \& Maeder, A. 2004, Astronomy \& Astrophysics, 425,
  649

\bibitem[{Howarth {et~al.}(1997)Howarth, Siebert, Hussain, \&
  Prinja}]{howarth97}
Howarth, I., Siebert, K., Hussain, G., \& Prinja, R. 1997, MNRAS, 284, 265

\bibitem[{Huang {et~al.}(2010)Huang, Gies, \& McSwain}]{huang2010}
Huang, W., Gies, D., \& McSwain, M.~V. 2010, The Astrophysical Journal, 722,
  605

\bibitem[{Huenemoerder {et~al.}(2012)Huenemoerder, Oskinova, Ignace, Waldron,
  Todt, Hamaguchi, \& Kitamoto}]{hue2012}
Huenemoerder, D.~P., Oskinova, L.~M., Ignace, R., {et~al.} 2012, The
  Astrophysical Journal Letters, 756, 34

\bibitem[{Hunter {et~al.}(2009)Hunter, Brott, Langer, Lennon, Dufton, Howarth,
  Ryans, Trundle, Evans, de~Koter, \& Smartt}]{hunter2009}
Hunter, I., Brott, I., Langer, N., {et~al.} 2009, Astronomy \& Astrophysics,
  496, 841

\bibitem[{Hunter {et~al.}(2008)Hunter, Brott, Lennon, Langer, Dufton, Trundle,
  Smartt, de~Koter, Evans, \& Ryans}]{hunter2008}
Hunter, I., Brott, I., Lennon, D.~J., {et~al.} 2008, The Astrophysical Journal,
  676, L29

\bibitem[{Iglesias \& Rogers(1993)}]{iglesias93}
Iglesias, C.~A. \& Rogers, F.~J. 1993, The Astrophysical Journal, 412, 752

\bibitem[{Iglesias \& Rogers(1996)}]{iglesias96}
Iglesias, C.~A. \& Rogers, F.~J. 1996, The Astrophysical Journal, 464, 943

\bibitem[{Jones {et~al.}(2015)Jones, Hirschi, Pignatari, Heger, Georgy,
  Nishimura, Fryer, \& Herwig}]{jones2015}
Jones, S., Hirschi, R., Pignatari, M., {et~al.} 2015, MNRAS, 447, 3115

\bibitem[{Kippenhahn {et~al.}(2012)Kippenhahn, Weigert, \&
  Weiss}]{kippenhahn2012}
Kippenhahn, R., Weigert, A., \& Weiss, A. 2012, Stellar Structure and Evolution
  (Springer-Verlag Berlin Heidelberg, 2nd edition)

\bibitem[{K{\"o}hler {et~al.}(2015)K{\"o}hler, Langer, de~Koter, de~Mink,
  Crowther, Evans, Gr{\"a}fener, Sana, Sanyal, Schneider, \& Vink}]{kohler2015}
K{\"o}hler, K., Langer, N., de~Koter, A., {et~al.} 2015, Astronomy \&
  Astrophysics, 573, 71

\bibitem[{Kudritzki {et~al.}(1995)Kudritzki, Lennon, \& Puls}]{kudritzki95}
Kudritzki, R.~P., Lennon, D.~J., \& Puls, J. 1995, in ESO Astrophysics
  Symposia, Science with the VLT, eds. J.R. Walsh and I.J. Danziger, Springer,
  Heidelberg, 46

\bibitem[{Kudritzki {et~al.}(1989)Kudritzki, Pauldrach, Puls, \&
  Abbott}]{kudritzki89}
Kudritzki, R.~P., Pauldrach, A.~W.~A., Puls, J., \& Abbott, D.~C. 1989,
  Astronomy \& Astrophysics, 219, 205

\bibitem[{Kudritzki \& Puls(2000)}]{kudritzki2000}
Kudritzki, R.~P. \& Puls, J. 2000, Annual Review of Astronomy \& Astrophysics,
  38, 613

\bibitem[{Lamers {et~al.}(1995)Lamers, Snow, \& Lindholm}]{lamers95}
Lamers, H.~J.~G.~L.~M., Snow, T.~P., \& Lindholm, D.~M. 1995, The Astrophysical
  Journal, 455, 269

\bibitem[{Langer(1986)}]{langer86}
Langer, N. 1986, Astronomy \& Astrophysics, 164, 45

\bibitem[{Langer(1998)}]{langer98}
Langer, N. 1998, Astronomy \& Astrophysics, 329, 551

\bibitem[{Langer {et~al.}(1988)Langer, Kiriakidis, El~Eid, Fricke, \&
  Weiss}]{langer88}
Langer, N., Kiriakidis, M., El~Eid, M.~F., Fricke, K.~J., \& Weiss, A. 1988,
  Astronomy \& Astrophysics, 192, 177

\bibitem[{Leutenegger {et~al.}(2013)Leutenegger, Cohen, Sundqvist, \&
  Owocki}]{leutenegger2013}
Leutenegger, M.~A., Cohen, D.~H., Sundqvist, J.~O., \& Owocki, S.~P. 2013, The
  Astrophysical Journal, 770, 80

\bibitem[{Lodders(2003)}]{lodders2003}
Lodders, K. 2003, The Astrophysical Journal, 591, 1220

\bibitem[{Lucy \& Solomon(1970)}]{lucy70}
Lucy, L.~B. \& Solomon, P.~M. 1970, The Astrophysical Journal, 159, 879

\bibitem[{{Maeder}(1999)}]{Maeder99}
{Maeder}, A. 1999, \aap, 347, 185

\bibitem[{Maeder(2009)}]{maeder2009}
Maeder, A. 2009, Physics, Formation and Evolution of Rotating Stars
  (Springer-Verlag Berlin Heidelberg)

\bibitem[{Maeder \& Meynet(2000)}]{maeder2000b}
Maeder, A. \& Meynet, G. 2000, Astronomy \& Astrophysics, 361, 159

\bibitem[{Maeder \& Meynet(2003)}]{maeder2003}
Maeder, A. \& Meynet, G. 2003, Astronomy \& Astrophysics, 411, 543

\bibitem[{Maeder \& Meynet(2005)}]{maeder2005}
Maeder, A. \& Meynet, G. 2005, Astronomy \& Astrophysics, 440, 1041

\bibitem[{Maeder \& Zahn(1998)}]{maeder98}
Maeder, A. \& Zahn, J.-P. 1998, Astronomy \& Astrophysics, 334, 1000

\bibitem[{Markova \& Puls(2008)}]{markova2008}
Markova, N. \& Puls, J. 2008, Astronomy \& Astrophysics, 478, 823

\bibitem[{Markova {et~al.}(2004)Markova, Puls, Repolust, \&
  Markov}]{markova2004}
Markova, N., Puls, J., Repolust, T., \& Markov, H. 2004, Astronomy \&
  Astrophysics, 413, 693

\bibitem[{{Martins} {et~al.}(2015{\natexlab{a}}){Martins}, {Herv{\'e}},
  {Bouret}, {Marcolino}, {Wade}, {Neiner}, {Alecian}, {Grunhut}, \&
  {Petit}}]{martins2015a}
{Martins}, F., {Herv{\'e}}, A., {Bouret}, J.-C., {et~al.} 2015{\natexlab{a}},
  \aap, 575, A34

\bibitem[{Martins \& Palacios(2013)}]{martins2013}
Martins, F. \& Palacios, A. 2013, Astronomy \& Astrophysics, 560, 16

\bibitem[{Martins {et~al.}(2005)Martins, Schaerer, \& Hillier}]{martins2005}
Martins, F., Schaerer, D., \& Hillier, D.~J. 2005, Astronomy \& Astrophysics,
  436, 1049

\bibitem[{{Martins} {et~al.}(2015{\natexlab{b}}){Martins},
  {Sim{\'o}n-D{\'{\i}}az}, {Palacios}, {Howarth}, {Georgy}, {Walborn},
  {Bouret}, \& {Barb{\'a}}}]{martins2015b}
{Martins}, F., {Sim{\'o}n-D{\'{\i}}az}, S., {Palacios}, A., {et~al.}
  2015{\natexlab{b}}, \aap, 578, A109

\bibitem[{McEvoy {et~al.}(2015)McEvoy, Dufton, Evans, Kalari, Markova,
  Sim{\'o}n-D{\'i}az, Vink, Walborn, Crowther, de~Koter, de~Mink, Dunstall,
  Henault-Brunet, Herrero, Langer, Lennon, Ma{\'i}z-Apell{\'a}niz, Najarro,
  Puls, Sana, Schneider, \& Taylor}]{mcevoy2015}
McEvoy, C.~M., Dufton, P.~L., Evans, C.~J., {et~al.} 2015, Astronomy \&
  Astrophysics, 575, 70

\bibitem[{Meynet {et~al.}(2011)Meynet, Eggenberger, \& Maeder}]{meynet2011}
Meynet, G., Eggenberger, P., \& Maeder, A. 2011, Astronomy \& Astrophysics,
  525, 11

\bibitem[{Meynet {et~al.}(2015)Meynet, Kudritzki, \& Georgy}]{meynet2015}
Meynet, G., Kudritzki, R.~P., \& Georgy, C. 2015, Astronomy \& Astrophysics
  (accepted for publication)

\bibitem[{{Meynet} {et~al.}(1994){Meynet}, {Maeder}, {Schaller}, {Schaerer}, \&
  {Charbonnel}}]{Meynet1994}
{Meynet}, G., {Maeder}, A., {Schaller}, G., {Schaerer}, D., \& {Charbonnel}, C.
  1994, \aaps, 103

\bibitem[{Mokiem {et~al.}(2007{\natexlab{a}})Mokiem, de~Koter, Evans, Puls,
  Smartt, Crowther, Herrero, Langer, Lennon, Najarro, Villamariz, \&
  Vink}]{mokiem2007b}
Mokiem, M.~R., de~Koter, A., Evans, C.~J., {et~al.} 2007{\natexlab{a}},
  Astronomy \& Astrophysics, 465, 1003

\bibitem[{Mokiem {et~al.}(2005)Mokiem, de~Koter, Puls, Herrero, Najarro, \&
  Villamariz}]{mokiem2005}
Mokiem, M.~R., de~Koter, A., Puls, J., {et~al.} 2005, Astronomy \&
  Astrophysics, 441, 711

\bibitem[{Mokiem {et~al.}(2007{\natexlab{b}})Mokiem, de~Koter, Vink, Puls,
  Evans, Smartt, Crowther, Herrero, Langer, Lennon, Najarro, \&
  Villamariz}]{mokiem2007a}
Mokiem, M.~R., de~Koter, A., Vink, J.~S., {et~al.} 2007{\natexlab{b}},
  Astronomy \& Astrophysics, 473, 603

\bibitem[{Moravveji {et~al.}(2015)Moravveji, Aerts, Papics, Triana, \&
  Vandoren}]{moravveji2015}
Moravveji, E., Aerts, C., Papics, P.~I., Triana, S.~A., \& Vandoren, B. 2015,
  Astronomy \& Astrophysics, 580, 14

\bibitem[{{M{\"u}ller} \& {Vink}(2014)}]{mueller2014}
{M{\"u}ller}, P.~E. \& {Vink}, J.~S. 2014, \aap, 564, A57

\bibitem[{Najarro {et~al.}(2011)Najarro, Hanson, \& Puls}]{najarro2011}
Najarro, F., Hanson, M.~M., \& Puls, J. 2011, Astronomy \& Astrophysics, 535,
  25

\bibitem[{Neiner {et~al.}(2015)Neiner, Mathis, Alecian, Emeriau, Grunhut, \&
  the BinaMIcS}]{neiner2015}
Neiner, C., Mathis, S., Alecian, E., {et~al.} 2015, Proceedings of the
  International Astronomical Union, 305, 61

\bibitem[{Nieuwenhuijzen \& de~Jager(1990)}]{nie90}
Nieuwenhuijzen, H. \& de~Jager, C. 1990, Astronomy \& Astrophysics, 231, 134

\bibitem[{{Oskinova} {et~al.}(2007){Oskinova}, {Hamann}, \&
  {Feldmeier}}]{oskinova07}
{Oskinova}, L.~M., {Hamann}, W.-R., \& {Feldmeier}, A. 2007, \aap, 476, 1331

\bibitem[{Pauldrach \& Puls(1990)}]{pauldrach90}
Pauldrach, A.~W.~A. \& Puls, J. 1990, Astronomy \& Astrophysics, 237, 409

\bibitem[{Pauldrach {et~al.}(1986)Pauldrach, Puls, \& Kudritzki}]{pauldrach86}
Pauldrach, A.~W.~A., Puls, J., \& Kudritzki, R.~P. 1986, Astronomy \&
  Astrophysics, 164, 86

\bibitem[{Paxton {et~al.}(2011)Paxton, Bildsten, Dotter, Herwig, Lesaffre, \&
  Timmes}]{paxton2011}
Paxton, B., Bildsten, L., Dotter, A., {et~al.} 2011, The Astrophysical Journal
  Supplement Series, 192, 3

\bibitem[{Paxton {et~al.}(2013)Paxton, Cantiello, Arras, Bildsten, Brown,
  Dotter, Mankovich, Montgomery, Stello, Timmes, \& Townsend}]{paxton2013}
Paxton, B., Cantiello, M., Arras, P., {et~al.} 2013, Astronomy \& Astrophysics
  Supplement Series, 208, 4

\bibitem[{Paxton {et~al.}(2015)Paxton, Marchant, Schwab, Bauer, Bildsten,
  Cantiello, Dessar, Farmer, Langer, Townsley, \& Timmes}]{paxton2015}
Paxton, B., Marchant, P., Schwab, J., {et~al.} 2015, Astronomy \& Astrophysics
  Supplement Series, 2, 4

\bibitem[{Petermann {et~al.}(2015)Petermann, Langer, Castro, \&
  Fossati}]{petermann2015}
Petermann, I., Langer, N., Castro, N., \& Fossati, L. 2015, Astronomy \&
  Astrophysics, 584, 7

\bibitem[{Petrov {et~al.}(2016)Petrov, Vink, \& Gr{\"a}fener}]{petrov2016}
Petrov, B., Vink, J., \& Gr{\"a}fener, G. 2016, MNRAS, 458, 1999

\bibitem[{Petrovic {et~al.}(2005)Petrovic, Langer, Yoon, \&
  Heger}]{petrovic2005}
Petrovic, J., Langer, N., Yoon, S.-C., \& Heger, A. 2005, Astronomy \&
  Astrophysics Supplement Series, 435, 247

\bibitem[{Potter {et~al.}(2012)Potter, Tout, \& Edridge}]{potter2012}
Potter, A.~T., Tout, C.~A., \& Edridge, J.~J. 2012, MNRAS, 419, 748

\bibitem[{Prinja {et~al.}(1990)Prinja, Barlow, \& Howarth}]{prinja90}
Prinja, R.~K., Barlow, M.~J., \& Howarth, I.~D. 1990, The Astrophysical
  Journal, 361, 607

\bibitem[{Prinja \& Massa(1998)}]{prinja98}
Prinja, R.~K. \& Massa, D.~L. 1998, in Boulder-Munich II: Properties of Hot,
  Luminous Stars, edited by Ian Howarth, Vol. 131, 218

\bibitem[{{Przybilla} {et~al.}(2010){Przybilla}, {Firnstein}, {Nieva},
  {Meynet}, \& {Maeder}}]{przybilla2010}
{Przybilla}, N., {Firnstein}, M., {Nieva}, M.~F., {Meynet}, G., \& {Maeder}, A.
  2010, \aap, 517, A38

\bibitem[{Puls {et~al.}(1996)Puls, Kudritzki, Herrero, Pauldrach, Haser,
  Lennon, Gabler, Voels, Vilchez, Wachter, \& Feldmeier}]{puls96}
Puls, J., Kudritzki, R.-P., Herrero, A., {et~al.} 1996, Astronomy \&
  Astrophysics, 305, 171

\bibitem[{Puls {et~al.}(2006)Puls, Markova, Scuderi, Stanghellini, Taranova,
  Burnley, \& Howarth}]{puls2006}
Puls, J., Markova, N., Scuderi, S., {et~al.} 2006, Astronomy \& Astrophysics,
  454, 625

\bibitem[{Puls {et~al.}(2000)Puls, Springmann, \& Lennon}]{puls2000}
Puls, J., Springmann, U., \& Lennon, M. 2000, Astronomy \& Astrophysics, 141,
  23

\bibitem[{Puls {et~al.}(2015)Puls, Sundqvist, \& Markova}]{puls2015}
Puls, J., Sundqvist, J.~O., \& Markova, N. 2015, in New windows on massive
  stars: asteroseismology, interferometry, and spectropolarimetry, Vol. 307,
  Proceedings of the International Astronomical Union, IAU Symposium, 25--36

\bibitem[{Puls {et~al.}(2008)Puls, Vink, \& Najarro}]{puls2008}
Puls, J., Vink, J.~S., \& Najarro, F. 2008, Annual Review of Astronomy \&
  Astrophysics, 16, 209

\bibitem[{Rauw {et~al.}(2015)Rauw, Herv\'e, Naz\'e, Gonz\'alez-P\'erez,
  Hempelmann, Mittag, Schmitt, Schr\:oder, Gosset, Eenens, \&
  Uuh-Sonda}]{rauw2015}
Rauw, G., Herv\'e, A., Naz\'e, Y., {et~al.} 2015, Astronomy \& Astrophysics,
  580, 16

\bibitem[{Repolust {et~al.}(2004)Repolust, Puls, \& Herrero}]{rep2004}
Repolust, T., Puls, J., \& Herrero, A. 2004, Astronomy \& Astrophysics, 414,
  349

\bibitem[{Rivero~Gonz{\'a}lez {et~al.}(2012)Rivero~Gonz{\'a}lez, Puls, Najarro,
  \& Brott}]{rivero2012}
Rivero~Gonz{\'a}lez, J.~G., Puls, J., Najarro, F., \& Brott, I. 2012, Astronomy
  \& Astrophysics, 537, A79

\bibitem[{R{\"u}diger {et~al.}(2012)R{\"u}diger, Kitchatinov, \&
  Elstner}]{ruediger2012}
R{\"u}diger, G., Kitchatinov, L.~L., \& Elstner, D. 2012, MNRAS, 425, 2267

\bibitem[{Schaller {et~al.}(1992)Schaller, Schaerer, Meynet, \&
  Maeder}]{schaller92}
Schaller, G., Schaerer, D., Meynet, G., \& Maeder, A. 1992, Astronomy \&
  Astrophysics Supplement Series, 96, 269

\bibitem[{Sim{\'o}n-D{\'i}az \& Herrero(2014)}]{simon2014}
Sim{\'o}n-D{\'i}az, S. \& Herrero, A. 2014, Astronomy \& Astrophysics
  Supplement Series, 562, A135

\bibitem[{Song {et~al.}(2016)Song, Meynet, Maeder, Ekstr{\:o}m, \&
  Eggenberger}]{song2016}
Song, H.~F., Meynet, G., Maeder, A., Ekstr{\:o}m, S., \& Eggenberger, P. 2016,
  Astronomy \& Astrophysics, 585, 21

\bibitem[{Spruit(2002)}]{spruit2002}
Spruit, H.~C. 2002, Astronomy \& Astrophysics Supplement Series, 381, 923

\bibitem[{Stello {et~al.}(2016)Stello, Cantiello, Fuller, Huber, Garcia,
  Bedding, Bildsten, \& Silva~Aguirre}]{stello2016}
Stello, D., Cantiello, M., Fuller, J., {et~al.} 2016, Nature, published online
  on Jan 4, 2016

\bibitem[{Sundqvist(2013)}]{sundqvist2013c}
Sundqvist, J.~O. 2013, Massive Stars: From alpha to Omega, held 10-14 June 2013
  in Rhodes, Greece, 47

\bibitem[{Sundqvist {et~al.}(2011)Sundqvist, Puls, Feldmeier, \&
  Owocki}]{sundqvist2011}
Sundqvist, J.~O., Puls, J., Feldmeier, A., \& Owocki, S.~P. 2011, Astronony \&
  Astrophysics, 528, 16

\bibitem[{Tayler(1973)}]{tayler73}
Tayler, R.~J. 1973, MNRAS, 161, 365

\bibitem[{{{\v S}urlan} {et~al.}(2013){{\v S}urlan}, {Hamann}, {Aret},
  {Kub{\'a}t}, {Oskinova}, \& {Torres}}]{Surlan13}
{{\v S}urlan}, B., {Hamann}, W.-R., {Aret}, A., {et~al.} 2013, \aap, 559, A130

\bibitem[{Vink {et~al.}(2001)Vink, de~Koter, \& Lamers}]{vink2001}
Vink, J., de~Koter, A., \& Lamers, H.~J.~G.~L.~M. 2001, Astronomy \&
  Astrophysics, 369, 574

\bibitem[{Vink {et~al.}(2010)Vink, Brott, Gr{\"a}fener, Langer, de~Koter, \&
  Lennon}]{vink2010}
Vink, J.~S., Brott, I., Gr{\"a}fener, G., {et~al.} 2010, Astronomy \&
  Astrophysics, 512, L7

\bibitem[{Vink {et~al.}(1999)Vink, de~Koter, \& Lamers}]{vink99}
Vink, J.~S., de~Koter, A., \& Lamers, H.~J.~G.~L.~M. 1999, Astronomy \&
  Astrophysics, 350, 181

\bibitem[{Vink {et~al.}(2000)Vink, de~Koter, \& Lamers}]{vink2000}
Vink, J.~S., de~Koter, A., \& Lamers, H.~J.~G.~L.~M. 2000, Astronomy \&
  Astrophysics, 362, 295

\bibitem[{Wade {et~al.}(2014)Wade, Grunhut, Alecian, Neiner, Auriere,
  Bohlender, David-Uraz, Folsom, Henrichs, Kochukhov, Mathis, Owocki, \&
  Petit}]{wade2014}
Wade, G.~A., Grunhut, J., Alecian, E., {et~al.} 2014, in Magnetic Fields
  throughout Stellar Evolution, Vol. 302, Proceedings of the International
  Astronomical Union, IAU Symposium, 265--269

\bibitem[{{Wade} \& {MiMeS Collaboration}(2015)}]{wade2015}
{Wade}, G.~A. \& {MiMeS Collaboration}. 2015, in Astronomical Society of the
  Pacific Conference Series, Vol. 494, Physics and Evolution of Magnetic and
  Related Stars, ed. Y.~Y. {Balega}, I.~I. {Romanyuk}, \& D.~O. {Kudryavtsev},
  30

\bibitem[{Yoon \& Langer(2005)}]{yoon2005}
Yoon, S.-C. \& Langer, N. 2005, Astronomy \& Astrophysics, 443, 643

\bibitem[{Zahn(1992)}]{zahn92}
Zahn, J.-P. 1992, Astronomy \& Astrophysics, 265, 115

\bibitem[{Zahn {et~al.}(2007)Zahn, Brun, \& Mathis}]{zahn2007}
Zahn, J.-P., Brun, A.~S., \& Mathis, S. 2007, Astronomy \& Astrophysics, 474,
  145

\end{thebibliography}

\end{document}